\documentclass[american]{article}
\usepackage[T1]{fontenc}
\usepackage[latin9]{inputenc}
\usepackage[active]{srcltx}
\usepackage{babel}
\usepackage{booktabs}
\usepackage{url}
\usepackage{amsmath}
\usepackage{amsthm}
\usepackage{amssymb}
\usepackage{graphicx}
\usepackage{wasysym}
\usepackage[unicode=true,
 bookmarks=true,bookmarksnumbered=false,bookmarksopen=false,
 breaklinks=false,pdfborder={0 0 1},backref=false,colorlinks=false]
 {hyperref}

\makeatletter

\providecommand{\tabularnewline}{\\}

\numberwithin{equation}{section}
\numberwithin{figure}{section}
\theoremstyle{plain}
\newtheorem{thm}{\protect\theoremname}
\theoremstyle{plain}
\newtheorem{prop}[thm]{\protect\propositionname}
\theoremstyle{remark}
\newtheorem{rem}[thm]{\protect\remarkname}

\usepackage{babel}
\usepackage{fullpage}
\usepackage{authblk}
\usepackage{hyperref}
\usepackage{placeins}
\usepackage{hyphenat}
\usepackage{fancyvrb}
\usepackage[hypcap]{caption}
\usepackage{bbm} 
\usepackage{relsize}
\usepackage{hyperref}
\usepackage{amsbsy}
\usepackage{textcomp}
\usepackage{pgf} 
\newcommand{\pt}[1]{\pgfmathprintnumber[fixed zerofill, precision=3]{#1} }
\newcommand{\ind}{ \mathlarger{\mathlarger{\mathbbm{1}}}}
\title{Robust Pricing of Equity-Indexed Annuities under Uncertain Volatility and Stochastic Interest Rate}
\author{ \textsc{Ludovic Goudenege}\thanks{Laboratoire de Mathématiques et Modélisation d'Évry (LaMME), CNRS UMR 8071, Université Paris-Saclay Évry, France - \texttt{ludovic.goudenege@math.cnrs.fr}} 
\and \textsc{Andrea Molent}\thanks{Dipartimento di Scienze Economiche e Statistiche, Universit\`a degli Studi di Udine, Italy - \texttt{andrea.molent@uniud.it}} 
\and \textsc{Antonino Zanette}\thanks{Dipartimento di Scienze Economiche e Statistiche, Universit\`a degli Studi di Udine, Italy - \texttt{antonino.zanette@uniud.it}}}
\date{}

\@ifundefined{showcaptionsetup}{}{%
 \PassOptionsToPackage{caption=false}{subfig}}
\usepackage{subfig}
\makeatother

\providecommand{\propositionname}{Proposition}
\providecommand{\remarkname}{Remark}
\providecommand{\theoremname}{Theorem}

\begin{document}
\maketitle

\begin{flushleft}
\rule{1\columnwidth}{1pt}
\par\end{flushleft}

\begin{flushleft}
\textbf{\large{}Abstract}{\large\par}
\par\end{flushleft}

In this paper, we propose a novel methodology for pricing equity-indexed
annuities featuring cliquet-style payoff structures and early surrender
risk, using advanced financial modeling techniques. Specifically,
the market is modeled by an equity index that follows an uncertain
volatility framework, while the dynamics of the interest rate are
captured by the Hull-White model. Due to the inherent complexity of
the market dynamics under consideration, we develop a numerical algorithm
that employs a tree-based framework to discretize both the interest
rate and the underlying equity index, enhanced with local volatility
optimization. The proposed algorithm is compared with a machine learning-based
algorithm. Extensive numerical experiments demonstrate its high effectiveness.
Furthermore, the numerical framework is employed to analyze key features
of the insurance contract, including the delineation of the optimal
exercise region when early surrender risk is incorporated.

\vspace{2mm}

\noindent \emph{\large{}Keywords}: Equity-indexed Annuities; Uncertain
Volatility Model; Stochastic interest Rate; Tree Method; Machine Learning.

\noindent\rule{1\columnwidth}{1pt}

\newpage

\section{Introduction}

Equity-linked annuities (ELAs) are significant contracts in the insurance
market, offering interest credits and benefit payouts linked to the
performance of one or more equity indices, such as the S\&P 500. Among
these, equity-indexed annuities (EIAs) stand out for their unique
combination of features: they guarantee a minimum rate of return while
also allowing policyholders to benefit from additional earnings tied
to the performance of the associated indices. These instruments provide
a compelling blend of market-linked growth potential and downside
protection. The guaranteed minimum interest rate ensures that policyholders
retain a baseline value for their investment, even in unfavorable
market conditions. This combination of security and growth potential
has driven the popularity of EIAs, particularly among individuals
focused on retirement planning, who seek a prudent balance between
risk and reward. The payoffs of the most common EIA contracts follow
a cliquet-style structure, meaning they are determined through a series
of periodic resets, with gains secured at each reset point. Such a
feature is also known as ratchet, symbolizing the mechanism by which
the accrued value can only move upward or remain fixed, never decreasing. 

In the early studies on this topic, EIAs were priced under the assumption
that the underlying equity index follows a geometric Brownian motion
with a constant interest rate, in line with standard financial modeling
approaches, as discussed by Gerber and Shiu \cite{gerber2003pricing}
and Hardy \cite{hardy2003investment}. However, given the typically
long maturities of EIAs, it is more appropriate to price these contracts
under a stochastic interest rate model rather than relying on a constant
rate. The Vasicek model has been one of the first stochastic interest
rate framework applied to EIA pricing. Specifically, Sheldon and Tan
\cite{sheldon2003valuation} use Monte Carlo simulations to derive
prices for various types of EIAs within the Vasicek framework. Building
on this, Kijima and Wong \cite{kijima2007pricing} develop a closed-form
pricing formula for ratchet EIAs with geometric index averaging under
the extended Vasicek model. They also introduce an efficient Monte
Carlo simulation algorithm for pricing EIAs with arithmetic index
averaging. Furthermore, as highlighted by Kijima and Wong \cite{kijima2007pricing},
the EIA pricing literature has largely overlooked early surrender
risk. Moreover, they identified addressing this oversight as a compelling
and open problem in the field. To this aim, Wei et al. \cite{wei2013pricing}
focus on improving the pricing of EIAs, particularly those with ratchet
features, minimum contract values and early surrender options.

Later, some authors have focused on describing the dynamics of the
underlying by considering stochastic volatility models and jump-diffusion
models. In this regard, Windcliff et al. \cite{windcliff2006numerical}
examine several numerical methods and volatility models for valuing
cliquet options, and they focus on the impacts of different volatility
assumptions, such as local volatility surfaces, jump-diffusion models,
and uncertain volatility models. More recently, Hieber \cite{hieber2017cliquet}
introduces a regime-switching Lévy model to value equity-linked life
insurance contracts with annual cliquet-style return guarantees and
incorporates macroeconomic regime changes via a Markov chain. Cui
et al. \cite{cui2017equity} develop an efficient method for pricing
ELAs with cliquet-style guarantees using advanced models such as regime-switching
Lévy processes and stochastic volatility models with jumps. Hess \cite{hess2018cliquet}
addresses the pricing of cliquet options within a jump-diffusion Lévy
framework. In this model, the underlying stock price process incorporates
both Brownian motion and compound Poisson jumps, capturing rare market
events such as crashes or sudden upward movements. 

It is widely recognized that cliquet options exhibit significant sensitivity
to unpredictable future changes in the volatility skew, as discussed
by Windcliff et al. \cite{windcliff2006numerical}, among others.
Relying on constant or local volatility models, such as Dupire's framework,
can result in substantial mispricing, necessitating more sophisticated
approaches to volatility modeling. Stochastic volatility models like
Heston or Hull-White, as well as hybrid jump-diffusion models, typically
involve a challenging and sometimes unreliable calibration process.
Furthermore, the effectiveness of delta hedging strategies within
these models is often unclear, as noted by Hess \cite{hess2018cliquet}.
Unlike many other exotic derivatives, the valuation of a cliquet option
is particularly dependent on the choice of model used to calibrate
against vanilla option prices and a careful modeling for volatility
is particularly important for these products because of the non convex
nature of their payoff function. Given these complexities, several
experts in applied mathematical finance, such as Wilmott \cite{wilmott2002cliquet},
Windcliff et al. \cite{windcliff2006numerical}, Hénaff and Martini
\cite{henaff2010model}, have advocated for the use of the Uncertain
Volatility Model (UVM), which provides a robust framework for both
pricing and evaluating the volatility risk associated with structured
products containing embedded cliquet options. Specifically, the UVM,
introduced by Avellaneda et al. \cite{avellaneda1995pricing}, bounds
the volatility of the underlying within predefined limits. Such a
model ensures robust pricing even under significant market uncertainty,
making it highly applicable for exotic options and derivatives that
are sensitive to volatility fluctuations.

Recently, researchers have increasingly explored the integration of
the UVM with stochastic interest rate frameworks in the valuation
of financial derivative instruments. The combination of these two
models allows for a robust management of market uncertainties by addressing
the dual complexities of random volatility and the fluctuating dynamics
of interest rates. This synergy is particularly effective when traditional
stochastic processes struggle to fully capture the variability in
market conditions. The UVM, by keeping volatility within specified
bounds, enables robust pricing strategies even in the presence of
considerable market uncertainty. This makes it particularly effective
for exotic options and derivatives with high sensitivity to volatility
dynamics. Meanwhile, stochastic interest rate models, such as those
based on Hull-White or CIR (Cox-Ingersoll-Ross) processes, provide
a realistic representation of interest rate movements, which are crucial
for accurate pricing in fixed-income markets or derivatives tied to
rate-sensitive assets. For instance, the integration of these models
has been particularly useful in pricing cliquet options and equity-linked
products, where path dependency and market conditions introduce additional
layers of complexity. In this regard, Hölzermann \cite{holzermann2021hull,holzermann2024pricing}
examines the classical Hull-White model for interest rate term structures
while incorporating volatility uncertainty, and proposes a new approach
to pricing zero-coupon bonds and interest rate derivatives. Recently,
Wang \cite{wang2024pricing} addresses the problem of pricing European
call options under uncertainty in both volatility and the risk-free
interest rate, and he proposes a numerical method based on a partial
derivative equation and an interior penalty method to solve the discretized
optimization problem which arises in this context. This integrated
approach continues to gain traction as it bridges the gap between
theoretical modeling and practical market application, offering enhanced
tools for risk management and pricing under uncertainty. All of these
advancements have significant implications for financial engineering,
especially in stress-testing and hedging strategies for derivative
portfolios.

This paper explores the combined effect of an uncertain volatility
model and a stochastic interest rate on EIAs. Specifically, the proposed
stochastic model extends the UVM by allowing the interest rate to
follow a stochastic process. Under risk-neutral probability, we consider
a model in which the risk-free rate is described by the Hull-White
model, while the underlying equity index of the EIA contract evolves
according to a geometric Brownian motion with a growth rate equal
to the risk-free rate and an unknown but bounded volatility. This
modeling development enables a deeper analysis of EIAs and reveals
relationships that, to the best of our knowledge, have not been previously
reported in the literature. Moreover, the proposed robust stochastic
model contributes to the existing literature by examining the pricing
of these products when policyholders have the option to terminate
the contract before its stipulated maturity date. In particular, following
Wei et al. \cite{wei2013pricing}, we study these contracts with early
surrender option under uncertain volatility and stochastic interest
rate. The model is of particular interest because, as Smith observed
in \cite{smith2002american}, there are different ways of defining
the concept of option price, since the writer and the buyer, who are
trying to hedge against speculative risks, have different optimal
exercise strategies. These remarkable results are made possible by
the development of an efficient numerical pricing method specifically
designed for this model. The proposed pricing algorithm, termed Tree
UVHW, interprets the pricing of a derivative instrument as a local
optimization problem that can be solved by resorting to a tree structure
and extending it to the stochastic interest rate. In particular, Tree
UVHW uses a tree structure to spread the interest rate in addition
to a tree approach for the underlying. An optimization algorithm for
optimal volatility computation and interpolation to handle different
possible values of the underlying volatility are also included. The
proposed algorithm is tested against an adapted version of the GTU
algorithm introduced by Goudenège et al. \cite{goudenege2024leveraging},
which, in contrast, leverages machine learning techniques. Whenever
feasible, results from the Monte Carlo method are also used as a benchmark
for comparison. The results obtained highlight the high accuracy of
the Tree UVHW method and the reduced computational time.

The remainder of the paper is structured as follows. Section 2 introduces
the financial model that underpins our investigation. Section 3 presents
our novel tree-based numerical algorithm for option pricing within
the proposed stochastic framework. Section 4 details the structure
of the EIA contract, highlighting its cliquet-style payoff features
and the incorporation of early surrender risk. Section 5 discusses
the application of the proposed Tree UVHW algorithm to EIA pricing.
Section 6 explains how the GTU algorithm can be adapted for EIA valuation.
Finally, Section 7 provides an extensive set of numerical experiments
to validate our methodology and explore the characteristics of these
insurance contracts.

\section{The financial market model\label{sec:The-financial-model}}

Let $S=\left(S_{t}\right)_{t\in\left[0,T\right]}$ denote the value
of the underlying equity index, which evolves according to the uncertain
volatility model. Specifically, under a suitable risk-neutral probability
measure $\mathbb{Q}$, the dynamics of $S$ are governed by the stochastic
differential equation
\begin{equation}
dS_{t}=\left(r_{t}-\eta\right)S_{t}\,dt+\sigma\left(t,S_{t},r_{t}\right)S_{t}\,dW_{t}^{S},\label{sde_1d}
\end{equation}
 where $S_{0}=s_{0}>0$ is the initial spot price, $r_{t}$ is the
stochastic short rate, $\eta$ is the constant dividend yield, $\sigma\left(t,S_{t},r_{t}\right)$
is the time- and state-dependent volatility, and $W^{S}=\left(W_{t}^{S}\right)_{t\in\left[0,T\right]}$
is a standard Wiener process. In the UVM, the volatility $\sigma\left(t,S_{t},r_{t}\right)$
varies within known bounds: 
\begin{equation}
\sigma_{\min}\leq\sigma\left(t,S_{t},r_{t}\right)\leq\sigma_{\max},\label{vol_bounds}
\end{equation}
where $\sigma_{\min}$ and $\sigma_{\max}$ are positive constants.
For the sake of brevity, we will write $\sigma$ instead of $\sigma\left(t,S_{t},r_{t}\right)$,
thus omitting the dependency of volatility on time, the underlying
value, and the interest rate. 

Here, we assume that the short rate $r=\left(r_{t}\right)_{t\in\left[0,T\right]}$
evolves according to the Hull-White model under $\mathbb{Q}$: 
\begin{equation}
dr_{t}=\kappa\left(\theta(t)-r_{t}\right)\,dt+\omega\,dW_{t}^{r},\label{hw_model}
\end{equation}
 where $\kappa>0$ is the mean-reversion rate, $\theta(t)$ is a deterministic
function ensuring the model fits the term structure, $\omega>0$ is
the volatility of the short rate, and $W^{r}=\left(W_{t}^{r}\right)_{t\in\left[0,T\right]}$
is a standard Wiener process such that $d\left\langle W_{t}^{S},W_{t}^{r}\right\rangle =\rho dt$.
As is common practice (see e.g. Brigo and Mercurio \cite{brigo2001interest}),
we define the market instantaneous forward rate at time $t=0$ for
a maturity $T$ as
\begin{equation}
f^{M}\left(0,T\right)=-\frac{\partial\ln P^{M}\left(0,T\right)}{\partial T},\label{eq:fM_0T}
\end{equation}
 where $P^{M}(0,T)$ denotes the market price of a zero-coupon bond
at $t=0$ with maturity $T$. The following equation facilitates the
unique definition of the function $\theta_{t}$, thereby ensuring
that theoretical zero-coupon bond prices correspond to those observed
in the market: 
\[
\theta\left(t\right)=\frac{1}{\kappa}\frac{\partial f^{M}(0,t)}{\partial t}+f^{M}(0,t)+\frac{\sigma^{2}}{2\kappa^{2}}\left(1-e^{-2\kappa t}\right).
\]
 Hence, the market instantaneous forward rate curve is derived based
on the observed prices of market bonds. In practice, this curve is
often simplified by fitting a parametric function. The Nelson--Siegel--Svensson
model \cite{svensson1994estimating} is widely adopted by central
banks and financial practitioners for modeling the term structure
of interest rates. In this framework, the market instantaneous forward
rate at $t=0$ is expressed as: 
\begin{equation}
f^{M}(0,T)=\beta_{0}+\beta_{1}e^{-\frac{T}{\tau_{1}}}+T\frac{\beta_{2}}{\tau_{1}}e^{-\frac{T}{\tau_{1}}}+T\frac{\beta_{3}}{\tau_{2}}e^{-\frac{T}{\tau_{2}}},\label{eq:fM}
\end{equation}
 where $\beta_{0},\beta_{1},\beta_{2},\beta_{3},\tau_{1},$ and $\tau_{2}$
are parameters determined through market calibration. A major achievement
in this field states that the short rate process $r_{t}$ can be expressed
as:
\[
r_{t}=\omega R_{t}+\beta(t),
\]
 where $R_{t}$ is a Vasicek process that evolves according to the
following dynamics: 
\begin{equation}
dR_{t}=-\kappa R_{t}dt+dW_{t}^{r},\quad R_{0}=0,\label{eq:R}
\end{equation}
 and $\beta(t)$ is a time-dependent function given by:
\begin{equation}
\beta(t)=f^{M}(0,t)+\frac{\omega^{2}}{2\kappa^{2}}\left(1-e^{-\kappa t}\right)^{2}.\label{eq:beta}
\end{equation}

\vspace{4mm}

In this framework, the price $V\left(t,S_{t},r_{t}\right)$ of a financial
derivative with maturity $T$ and payoff function $\Psi(S_{T})$ can
be formulated as an optimization problem to account for volatility
uncertainty. In particular, both the writer and the buyer of a financial
derivative seek the worst-case volatility scenario to ensure robust
pricing. Since any agent who has entered a short position on the contract
fears the rise of the contract value, we specify the option price
from such a perspective as
\begin{equation}
V^{Short}\left(t,S_{t},r_{t}\right)=\sup_{\sigma\in[\sigma_{\min},\sigma_{\max}]}\mathbb{E}^{\mathbb{Q}}\left[e^{-\int_{t}^{T}r_{u}\,du}\Psi(S_{T})\mid\mathcal{F}_{t}\right],\label{pricing_problem}
\end{equation}
where $\mathbb{E}^{\mathbb{Q}}$ denotes the expectation under $\mathbb{Q}$
and $\mathcal{F}=\left(\mathcal{F}_{t}\right)_{t\in\left[0,T\right]}$
represents the filtration generated by the processes $S$ and $r$.
Conversely, an economic agent who has assumed a long position on the
derivative fears a drop in the contract price. For this reason, the
value of the derivative is expressed as:

\begin{equation}
V^{Long}\left(t,S_{t},r_{t}\right)=\inf_{\sigma\in[\sigma_{\min},\sigma_{\max}]}\mathbb{E}^{\mathbb{Q}}\left[e^{-\int_{t}^{T}r_{u}\,du}\Psi(S_{T})\mid\mathcal{F}_{t}\right].\label{pricing_problem-1}
\end{equation}

These two formulations (\ref{pricing_problem}) and (\ref{pricing_problem-1})
ensure that the derivative price reflects the most adverse volatility
scenario, yielding a conservative and robust valuation. The Hull-White
model for $r_{t}$ further introduces realistic interest rate dynamics,
making this approach suitable for environments with both stochastic
interest rates and uncertain volatility. 

As far as an American contract is considered, we can identify three
relevant formulations for the option price which take into account
that the optimal exercise strategy is determined solely by the option
buyer. Then, the price for the long agent is defined as follows:

\begin{equation}
V_{am}^{Long}\left(t,S_{t},r_{t}\right)=\sup_{\tau\in\mathcal{T}}\inf_{\sigma\in[\sigma_{\min},\sigma_{\max}]}\mathbb{E}^{\mathbb{Q}}\left[\exp\left(-\int_{t}^{\tau}r_{u}\,du\right)\Psi(S_{\tau})\mid\mathcal{F}_{t}\right],\label{am1}
\end{equation}
where $\mathcal{T}$ is the set of all stopping times with respect
to the filtration $\mathcal{F}$. Let $\tau_{\star}$ be the optimal
stopping time, solving problem (\ref{am1}). As far as the short agent
is considered, the price can be defined as follows, based on the stopping
strategy $\tau_{\star}$ followed by the buyer:

\begin{equation}
V_{am}^{Mix}\left(t,S_{t},r_{t}\right)=\sup_{\sigma\in[\sigma_{\min},\sigma_{\max}]}\mathbb{E}^{\mathbb{Q}}\left[\exp\left(-\int_{t}^{\tau_{\star}}r_{u}\,du\right)\Psi(S_{\tau_{\star}})\mid\mathcal{F}_{t}\right],\label{am1-1}
\end{equation}

Finally, the writer may be interested in the maximum hedging cost
of the option, which can be computed by assuming that the buyer does
not choose the strategy that is most convenient for him but the strategy
that is most expensive for the writer. This leads to the following
formulation of the maximum option price for the short agent, which
he should take into account for precautionary purposes, in case the
buyer does not follow strategy $\tau_{\star}$: 

\begin{equation}
V_{am}^{Short}\left(t,S_{t},r_{t}\right)=\sup_{\tau\in\mathcal{T}}\sup_{\sigma\in[\sigma_{\min},\sigma_{\max}]}\mathbb{E}^{\mathbb{Q}}\left[\exp\left(-\int_{t}^{\tau}r_{u}\,du\right)\Psi(S_{\tau})\mid\mathcal{F}_{t}\right].\label{am1-2}
\end{equation}

\section{\label{sec:4}The Tree UVHW method for option pricing}

This Section presents a general version of the Tree UVHW algorithm,
which can be applied to any European vanilla option. The algorithm
employs two structures to discretize the market: a tree $\mathcal{R}$
for the interest rate process $R$ and a grid of representative values
$\mathcal{S}$ for the underlying equity index $S$. The values and
transition probabilities are determined to match the first two moments
of $R$ and of $\ln\left(S\right)$, as well as their covariance.
It should be noted that the tree structure for $R$ is recombinant
in the sense that future values originating from a given node remain
within the tree structure. In contrast, the lattice for $S$ does
not exhibit this characteristic, and linear interpolation is used
to calculate conditional expectations at future time steps (a similar
technique has been used by Costabile et al. \cite{costabile2020evaluating}
in a different insurance context). Additionally, the grid's ability
to account for uncertainty in index volatility is advantageous, as
it enables the management of different volatility values. 

Before introducing the proposed pricing method in detail, we recall
some properties of the Black-Scholes (BS) model with stochastic interest
rate modeled by a Hull-White process. 
\begin{prop}
\label{prop:3}Let $R$ be the stochastic process defined in (\ref{eq:R})
and let $W^{S}$ be a Brownian motion as in (\ref{sde_1d}). Moreover,
let $I$ be the stochastic process representing the integral
\[
I_{t}=\int_{0}^{t}R_{s}\,ds.
\]
Then, for any time $t>0$ and time increment $\Delta t>0$, the following
relation holds:
\[
\left(\begin{array}{c}
R_{t+\Delta t}-R_{t}\\
I_{t+\Delta t}-I_{t}\\
W_{t+\Delta t}-W_{t}
\end{array}\right)\mid R_{t}\sim\mathcal{N}\left(\left(\begin{array}{c}
\mu_{R}\\
\mu_{I}\\
\mu_{W}
\end{array}\right),\left(\begin{array}{ccc}
\sigma_{R}^{2} & \sigma_{R,I} & \sigma_{R,W}\\
\sigma_{R,I} & \sigma_{I}^{2} & \sigma_{I,W}\\
\sigma_{R,W} & \sigma_{I,W} & \sigma_{W}^{2}
\end{array}\right)\right)
\]
with
\[
\mu_{R}=R_{t}\left(e^{-\kappa\Delta t}-1\right),\ \mu_{I}=-\frac{\mu_{R}}{\kappa},\ \mu_{W}=0,
\]
\[
\sigma_{R}^{2}=\frac{1}{2\kappa}\left(1-e^{-2\kappa\Delta t}\right),\sigma_{I}^{2}=\frac{1}{2\kappa^{3}}\left(2\kappa\Delta t+4e^{-\kappa\Delta t}-e^{-2\kappa\Delta t}-3\right),\sigma_{W}^{2}=\Delta t,
\]
\[
\sigma_{R,I}=\frac{1}{2\kappa^{2}}\left(1-e^{-\kappa\Delta t}\right)^{2},\sigma_{R,W}=\frac{\rho}{\kappa}\left(1-e^{-\kappa\Delta t}\right),\sigma_{I,W}=\frac{\rho}{\kappa^{2}}\left(\kappa\Delta t+e^{-\kappa\Delta t}-1\right).
\]
\end{prop}

The proof of this key result is reported in Appendix \ref{sec:Appendix_A}.

The Tree UVHW method for evaluating derivatives exploits lattice structures
to discretize the stochastic processes that govern both the interest
rate and the underlying asset index. This step involves dividing the
time interval between inception and maturity dates into $N_{T}$ sub-intervals
of equal length $\Delta t$. As a result, $N_{T}=T/\Delta t$ represents
the total number of time steps. Additionally, let $t_{n}=n\cdot\Delta t$
a generic time step. Let us denote $\bar{S}=\left\{ \bar{S}_{n}\right\} _{n=0,\dots,N_{T}}$
and $\bar{R}=\left\{ \bar{R}_{n}\right\} _{n=0,\dots,N_{T}}$ the
discrete time processes based used to approximate $S$ and $R$ respectively.
In particular, the couple $\left(\bar{S}_{n},\bar{R}_{n}\right)$
approximates the couple $\left(S_{t_{n}},R_{t_{n}}\right)$.

The first step of the proposed algorithm is the construction of the
tree structure for $\bar{R}$, which discretizes the interest rate
dynamics. At each time step, the tree branches into several potential
future states, capturing the possible evolution of the interest rate.
The associated rate process $\bar{R}$ at each tree node is updated
using a discretized version of the Hull-White model's stochastic differential
equation, ensuring that the tree remains recombining, meaning it merges
into fewer states over time to enhance computational efficiency. Specifically,
we employ the ``multiple-jumps'' tree introduced by Nelson and Ramaswamy
\cite{nelson1990simple} and further developed by Briani et al. in
\cite{briani2017hybrid}. The lattice structure is as follows: for
$n=0,1,\ldots,N_{T}$ we set
\[
\mathcal{R}_{n}=\{\bar{R}_{n,j}\}_{j=0,1,\ldots,n},\quad\text{with }\bar{R}_{n,j}=(2j-n)\sqrt{\Delta t}.
\]
For each fixed $R_{n,j}\in\mathcal{R}_{n}$, the ``\emph{up}'' and
``\emph{down}'' transitions value for $\bar{R}_{n}=\bar{R}_{n,j}$
are 
\[
\bar{R}_{n+1}^{up}\left(j\right)=\bar{R}_{n+1,j_{up}(n,j)}\text{ and }\bar{R}_{n+1}^{dw}\left(j\right)=\bar{R}_{n+1,j_{dw}(n,j)},
\]
respectively. We stress out that both $\bar{R}_{n+1}^{up}\left(j\right)$
and $\bar{R}_{n+1}^{dw}\left(j\right)$ are elements of the set $\mathcal{R}_{n+1}$.
The indices of these transitions, $j_{u}(n,j)$ and $j_{d}(n,j)$,
are defined as:

\[
\begin{aligned}j_{u}(n,j) & =\min\left\{ j^{*}\text{ s.t. }j+1\leq j^{*}\leq n+1\text{ and }\bar{R}_{n,j}+\mu_{R}\left(\bar{R}_{n,j}\right)\Delta t\leq\bar{R}_{n+1,j^{*}}\right\} ,\\
j_{d}(n,j) & =\max\left\{ j^{*}\text{ s.t. }0\leq j^{*}\leq j\text{ and }\bar{R}_{n,j}+\mu_{R}\left(\bar{R}_{n,j}\right)\Delta t\geq\bar{R}_{n+1,j^{*}}\right\} ,
\end{aligned}
\]
where $\mu_{R}\left(\bar{R}_{n,j}\right)=-\kappa\bar{R}_{n,j}$ represents
the drift of $R$. Furthermore, if at time step $n$ the process $\bar{R}$
takes the value corresponding to the node $(n,j)$, the probability
that the process $\bar{R}$ moves to $j_{u}\left(n,j\right)$ or to
$j_{d}\left(n,j\right)$ at time-step $n+1$ are given by
\[
p_{n,j}^{\bar{R},up}=0\vee\frac{\mu_{R}\left(\bar{R}_{n,j}\right)\Delta t+\bar{R}_{n,j}-\bar{R}_{n+1}^{dw}\left(j\right)}{\bar{R}_{n+1}^{up}\left(j\right)-\bar{R}_{n+1}^{dw}\left(j\right)}\wedge1\quad\text{ and }\quad p_{n,j}^{\bar{R},dw}=1-p_{n,j}^{\bar{R},up}.
\]

As far as the underlying is considered, we also exploit a lattice
to discretize the process $S$. Usually, lattice methods for pricing,
such as the CRR tree model \cite{cox1979crr} or the Jarrow Rudd model
\cite{jarrow1983option} require one to define the lattice based on
the volatility of $S$, which is not possible in our case, since in
UVM volatility is a single parameter, but a range of values $\left[\sigma_{\min},\sigma_{\max}\right]$.
In order to overcome this obstacle, the following approach is adopted.
Since, the greater the value for $\sigma$, the greater the variance
of the underlying, we define at each time step $t_{n}$ the extreme
values of the mesh for $S$ based on $\sigma_{\max}$ and on CRR model.
In particular, we set 
\[
\bar{S}_{n}^{\max}=S_{0}\exp\left(n\sigma_{\max}\sqrt{\Delta t}\right),\quad\bar{S}_{n}^{\min}=S_{0}\exp\left(-n\sigma_{\max}\sqrt{\Delta t}\right).
\]
We then define a uniform logarithmic mesh, denoted as $\mathcal{S}_{n}$,
consisting of points evenly distributed on a logarithmic scale between
$\bar{S}_{n}^{\min}$ and $\bar{S}_{n}^{\max}$. The number of points
used to define this mesh is an integer multiple of $n+1$ which corresponds
to the number of points at time-step $t_{n}$ that would be obtained
in the case of a binomial tree. Specifically, such a number is equal
to $N_{S}^{n}=N_{S}\left(n+1\right)$, with $N_{S}$ a large enough
integer, thus 
\[
\mathcal{S}_{n}=\left\{ \bar{S}_{n,j}=S_{0}\exp\left(\left[\frac{2n}{N_{S}\left(n+1\right)-1}\left(i-1\right)-n\right]\sigma_{\max}\sqrt{\Delta t}\right),\ i=1,\dots,N_{S}^{n}\right\} .
\]
Furthermore, since the starting value of the discrete process $\bar{S}$
coincides with that of the original continuous process $S$, we set
\[
\mathcal{S}_{0}=\left\{ S_{0}\right\} .
\]

Now, let us discuss how to define the transition dynamics of process
$\bar{S}$, that is how to discretize the random variable $\left(S_{t+\Delta t}\mid S_{t},R_{t},R_{t+\Delta t}\right)$.
We begin this discussion by stating the following result.
\begin{prop}
\label{prop:4}The random variable $\left(W_{t+\Delta t}^{S}-W_{t}^{S}\mid R_{t+\Delta t},R_{t}\right)$
has the following probability distribution:
\[
\left(W_{t+\Delta t}^{S}-W_{t}^{S}\mid R_{t+\Delta t},R_{t}\right)\sim\mathcal{N}\left(\mu_{W\mid R},\sigma_{W\mid R}^{2}\right),
\]
with

\[
\mu_{W\mid R}=2\rho\frac{1-e^{-\kappa\Delta t}}{1-e^{-2\kappa\Delta t}}\left(R_{t+\Delta t}-R_{t}e^{-\kappa\Delta t}\right),\quad\sigma_{W\mid R}^{2}=\Delta t-2\frac{\rho^{2}}{\kappa}\frac{1-e^{-\kappa\Delta t}}{1+e^{-\kappa\Delta t}}.
\]
\end{prop}

The proof us such a result is reported in Appendix \ref{sec:Appendix_B}. 

Now, let us focus on a particular market state, that is we consider
$\left(\bar{S}_{n},\bar{R}_{n}\right)=\left(\bar{S}_{n,i},\bar{R}_{n,j}\right)$.
The possible future values for the rate process $\bar{R}_{n+1}$ are
$\bar{R}_{n+1,j_{u}(n,j)}$ and $\bar{R}_{n+1,j_{d}(n,j)}$, respectively,
as described above. For the time being, let us assume that, during
the time interval $\left[t_{n},t_{n}+\Delta t\right]$, the volatility
of $\bar{S}$ is constant, equal to a specific value $\hat{\sigma}\in\left[\sigma_{\min},\sigma_{\max}\right]$.
Generally speaking, we can observe that, if $S_{t_{n}}=S_{n,i}$,
the solution of the SDE \ref{sde_1d} in $\left[t_{n},T\right]$ reads
out
\begin{equation}
S_{u}=S_{n,i}\exp\left(\int_{t_{n}}^{u}r_{s}ds-\left(\eta+\frac{\sigma^{2}}{2}\right)\left(u-t_{n}\right)+\sigma\left(W_{u}^{S}-W_{t_{n}}^{S}\right)\right)\label{eq:S_u}
\end{equation}
for any $u\in\left[t_{n},T\right]$. In particular, for $u=t_{n+1}$,
we can write down
\begin{equation}
S_{t_{n+1}}=S_{n,i}\exp\left(\int_{t_{n}}^{t_{n+1}}r_{s}ds-\left(\eta+\frac{\sigma^{2}}{2}\right)\Delta t+\sigma\left(W_{t_{n+1}}^{S}-W_{t_{n}}^{S}\right)\right).\label{eq:S_u-1}
\end{equation}
Therefore, to discretize the random value $S_{t_{n+1}}$ given the
future value $R_{t_{n+1}}$, that is $\bar{R}_{n+1}$, we develop
the following approximation of the integral, based on the trapezoidal
rule:
\[
\int_{t_{n}}^{t_{n+1}}r_{s}ds=\omega\int_{t_{n}}^{t_{n+1}}R_{s}ds+\int_{t_{n}}^{t_{n+1}}\beta\left(s\right)ds\approx\omega\frac{\bar{R}_{n+1}+\bar{R}_{n}}{2}\Delta t+\int_{t_{n}}^{t_{n+1}}\beta\left(s\right)ds.
\]
In this regard, note that the integral of the function $\beta$ can
be calculated exactly from the relations (\ref{eq:beta}) and (\ref{eq:fM}).

Furthermore, by exploiting Proposition \ref{prop:4}, we replace the
Gaussian random variable $W_{t_{n+1}}^{S}-W_{t_{n}}^{S}$ with a discrete
random variable $H_{n}$ having the same first two moments:
\[
H_{n}=\mu_{W\mid R}+\sqrt{\sigma_{W\mid R}}\left(2B-1\right),
\]
with $B\sim\mathcal{B}\left(0.5\right)$ a Bernoulli's random variable
independent from all the others. Moreover, let $H^{up}=\mu_{W\mid R}+\sqrt{\sigma_{W\mid R}}$
be the value for $H$ if $B=1$ and $H^{dw}=\mu_{W\mid R}-\sqrt{\sigma_{W\mid R}}$
the value for $H$ if $B=0$. Based on the two possible values for
$\bar{R}_{n+1}$, we replace the random variable $\left(W_{t_{n+1}}^{S}-W_{t}^{S}\mid\bar{R}_{t_{n+1}},\bar{R}_{t}\right)$
in (\ref{eq:S_u-1}) with $H_{n}$, so that there are four values
for $\left(\bar{S}_{n+1}\mid\bar{S}_{n}=\bar{S}_{n,i},\bar{R}_{n}=\bar{R}_{n,j},\bar{R}_{n+1}\right)$,
namely $\bar{S}_{n+1}^{up,up}\left(i,j\right)$, $\bar{S}_{n+1}^{up,up}\left(i,j\right),\bar{S}_{n+1}^{up,dw}\left(i,j\right)$
and $\bar{S}_{n+1}^{dw,dw}\left(i,j\right)$. Specifically, if $\bar{R}_{n+1}=\bar{R}_{n+1}^{up}\left(j\right)$,
then 
\begin{align}
\bar{S}_{n+1}^{up,up}\left(i,j\right) & =\bar{S}_{n,i}\exp\left(\omega\frac{\bar{R}_{n+1}^{up}\left(j\right)+\bar{R}_{n}}{2}\Delta t+\int_{t_{n}}^{t_{n+1}}\beta\left(s\right)ds-\left(\eta+\frac{\sigma^{2}}{2}\right)\Delta t+\sigma H^{up}\right),\label{eq:suu}\\
\bar{S}_{n+1}^{up,dw}\left(i,j\right) & =\bar{S}_{n,i}\exp\left(\omega\frac{\bar{R}_{n+1}^{up}\left(j\right)+\bar{R}_{n}}{2}\Delta t+\int_{t_{n}}^{t_{n+1}}\beta\left(s\right)ds-\left(\eta+\frac{\sigma^{2}}{2}\right)\Delta t+\sigma H^{dw}\right),\label{eq:sud}
\end{align}
while if $\bar{R}_{n+1}=\bar{R}_{n+1}^{dw}\left(j\right)$, then 
\begin{align}
\bar{S}_{n+1}^{dw,up}\left(i,j\right) & =\bar{S}_{n,i}\exp\left(\omega\frac{\bar{R}_{n+1}^{dw}\left(j\right)+\bar{R}_{n}}{2}\Delta t+\int_{t_{n}}^{t_{n+1}}\beta\left(s\right)ds-\left(\eta+\frac{\sigma^{2}}{2}\right)\Delta t+\sigma H^{up}\right),\label{eq:sdu}\\
\bar{S}_{n+1}^{dw,dw}\left(i,j\right) & =\bar{S}_{n,i}\exp\left(\omega\frac{\bar{R}_{n+1}^{dw}\left(j\right)+\bar{R}_{n}}{2}\Delta t+\int_{t_{n}}^{t_{n+1}}\beta\left(s\right)ds-\left(\eta+\frac{\sigma^{2}}{2}\right)\Delta t+\sigma H^{dw}\right).\label{eq:sdd}
\end{align}

Moreover, in agreement with the probability distribution of $B$,
the following equations hold:
\begin{multline*}
\mathbb{Q}\left(\bar{S}_{n+1}=\bar{S}_{n+1}^{up,up}\left(i,j\right)\mid\bar{S}_{n}=\bar{S}_{n,i},\bar{R}_{n}=\bar{R}_{n,j},\bar{R}_{n+1}=\bar{R}_{n+1}^{up}\left(j\right)\right)\\
=\mathbb{Q}\left(\bar{S}_{n+1}=\bar{S}_{n+1}^{up,dw}\left(i,j\right)\mid\bar{S}_{n}=\bar{S}_{n,i},\bar{R}_{n}=\bar{R}_{n,j},\bar{R}_{n+1}=\bar{R}_{n+1}^{up}\left(j\right)\right)=\frac{1}{2}
\end{multline*}
and 
\begin{multline*}
\mathbb{Q}\left(\bar{S}_{n+1}=\bar{S}_{n+1}^{dw,up}\left(i,j\right)\mid\bar{S}_{n}=\bar{S}_{n,i},\bar{R}_{n}=\bar{R}_{n,j},\bar{R}_{n+1}=\bar{R}_{n+1}^{dw}\left(j\right)\right)\\
=\mathbb{Q}\left(\bar{S}_{n+1}=\bar{S}_{n+1}^{dw,dw}\left(i,j\right)\mid\bar{S}_{n}=\bar{S}_{n,i},\bar{R}_{n}=\bar{R}_{n,j},\bar{R}_{n+1}=\bar{R}_{n+1}^{dw}\left(j\right)\right)=\frac{1}{2}
\end{multline*}
As opposed to the two values for $\bar{R}_{n+1}$ that belong to $\mathcal{R}_{n+1}$,
the four values for $\bar{S}_{n+1}$ may not belong to $\mathcal{S}_{n+1}$,
but the use of one dimensional linear interpolation on $\mathcal{S}_{n+1}$
allows one to overcome this obstacle. Specifically, one can use the
previous definitions to compute discounted conditional expectation
of a function $f$ defined over the grid $\mathcal{R}_{n+1}\times\mathcal{S}_{n+1}$.
To this aim, one has to extend $f$ to $\mathbb{R}^{2}$ by linear
interpolation on $\mathcal{R}_{n+1}\times\mathcal{S}_{n+1}$, and
then compute
\begin{multline}
\mathbb{E}\left[\exp\left(-\int_{t_{n}}^{t_{n+1}}\omega R_{s}+\beta\left(s\right)ds\right)f\left(R_{t_{n+1}},S_{t_{n+1}}\right)\mid R_{t_{n}}=\bar{R}_{n,j},S_{t_{n}}=\bar{S}_{n,i}\right]\approx\\
\frac{1}{2}p_{n,j}^{\bar{R},up}d_{n}^{up}\left(j\right)\left[f\left(\bar{R}_{n+1}^{up}\left(j\right),\bar{S}_{n+1}^{up,up}\left(i,j\right)\right)+f\left(\bar{R}_{n+1}^{up}\left(j\right),\bar{S}_{n+1}^{up,dw}\left(i,j\right)\right)\right]+\\
\frac{1}{2}p_{n,j}^{\bar{R},dw}d_{n}^{dw}\left(j\right)\left[f\left(\bar{R}_{n+1}^{dw}\left(j\right),\bar{S}_{n+1}^{dw,up}\left(i,j\right)\right)+f\left(\bar{R}_{n+1}^{dw}\left(j\right),\bar{S}_{n+1}^{dw,dw}\left(i,j\right)\right)\right],\label{eq:expected}
\end{multline}
with
\begin{align*}
d_{n}^{up}\left(j\right) & =\exp\left(-\omega\frac{\bar{R}_{n+1}^{up}\left(j\right)+\bar{R}_{n,j}}{2}\Delta t-\int_{t_{n}}^{t_{n+1}}\beta\left(s\right)ds\right),\\
d_{n}^{dw}\left(j\right) & =\exp\left(-\omega\frac{\bar{R}_{n+1}^{dw}\left(j\right)+\bar{R}_{n,j}}{2}\Delta t-\int_{t_{n}}^{t_{n+1}}\beta\left(s\right)ds\right),
\end{align*}
 the discount factors from $t_{n+1}$ to $t_{n}$ in case of an up
or down movement respectively. In particular, if we define 
\begin{align*}
\bar{i}{}^{up,up} & \left(i,j\right)=\min\left\{ i^{*}\text{ s.t. }1\leq i^{*}\leq N_{S}^{n+1}\text{ and }\bar{S}_{n+1,i^{*}}\geq\bar{S}_{n+1}^{up,up}\left(i,j\right)\right\} \cup\left\{ N_{S}^{n+1}\right\} ,\\
\underline{i}^{up,up} & \left(i,j\right)=\max\left\{ i^{*}\text{ s.t. }1\leq i^{*}\leq N_{S}^{n+1}\text{ and }\bar{S}_{n+1,i^{*}}\leq\bar{S}_{n+1}^{up,up}\left(i,j\right)\right\} \cup\left\{ 1\right\} ,
\end{align*}
then linear interpolation for $f\left(\bar{S}_{n+1}^{up,up}\left(i,j\right),\bar{R}_{n+1}^{up}\left(j\right)\right)$
writes down as
\begin{multline*}
f\left(\bar{R}_{n+1}^{up}\left(j\right),\bar{S}_{n+1}^{up,up}\left(i,j\right)\right)=\frac{\bar{S}_{n+1}^{up,up}\left(i,j\right)-\bar{S}_{n+1,\underline{i}^{up,up}\left(i,j\right)}}{\bar{S}_{n+1,\bar{i}{}^{up,up}}-\bar{S}_{n+1,\underline{i}^{up,up}\left(i,j\right)}}f\left(\bar{R}_{n+1}^{up}\left(j\right),\bar{S}_{n+1,\bar{i}{}^{up,up}\left(i,j\right)}\right)\\
+\frac{\bar{S}_{n+1,\bar{i}{}^{up,up}\left(i,j\right)}-\bar{S}_{n+1}^{up,up}\left(i,j\right)}{\bar{S}_{n+1,\bar{i}{}^{up,up}}-\bar{S}_{n+1,\underline{i}^{up,up}\left(i,j\right)}}f\left(\bar{R}_{n+1}^{up}\left(j\right),\bar{S}_{n+1,\underline{i}^{up,up}\left(i,j\right)}\right),
\end{multline*}
where $f\left(\bar{R}_{n+1}^{up}\left(j\right),\bar{S}_{n+1,\bar{i}{}^{up,up}\left(i,j\right)}\right)$
and $f\left(\bar{R}_{n+1}^{up}\left(j\right),\bar{S}_{n+1,\underline{i}^{up,up}\left(i,j\right)}\right)$
are observable values since $\bar{R}_{n+1}^{up}\left(j\right)\in\mathcal{R}_{n+1}$,
$\bar{S}_{n+1,\bar{i}{}^{up,up}}\in\mathcal{S}_{n+1}$ and $\bar{S}_{n+1,\underline{i}^{up,up}}\in\mathcal{S}_{n+1}$.
In a similar manner, it is possible to compute linear interpolation
for $f\left(\bar{R}_{n+1}^{up}\left(j\right),\bar{S}_{n+1}^{up,dw}\left(i,j\right)\right)$,
$f\left(\bar{R}_{n+1}^{dw}\left(j\right),\bar{S}_{n+1}^{dw,up}\left(i,j\right)\right)$
and $f\left(\bar{R}_{n+1}^{dw}\left(j\right),\bar{S}_{n+1}^{dw,dw}\left(i,j\right)\right)$.

\begin{figure}
\begin{centering}
\includegraphics[width=1\textwidth]{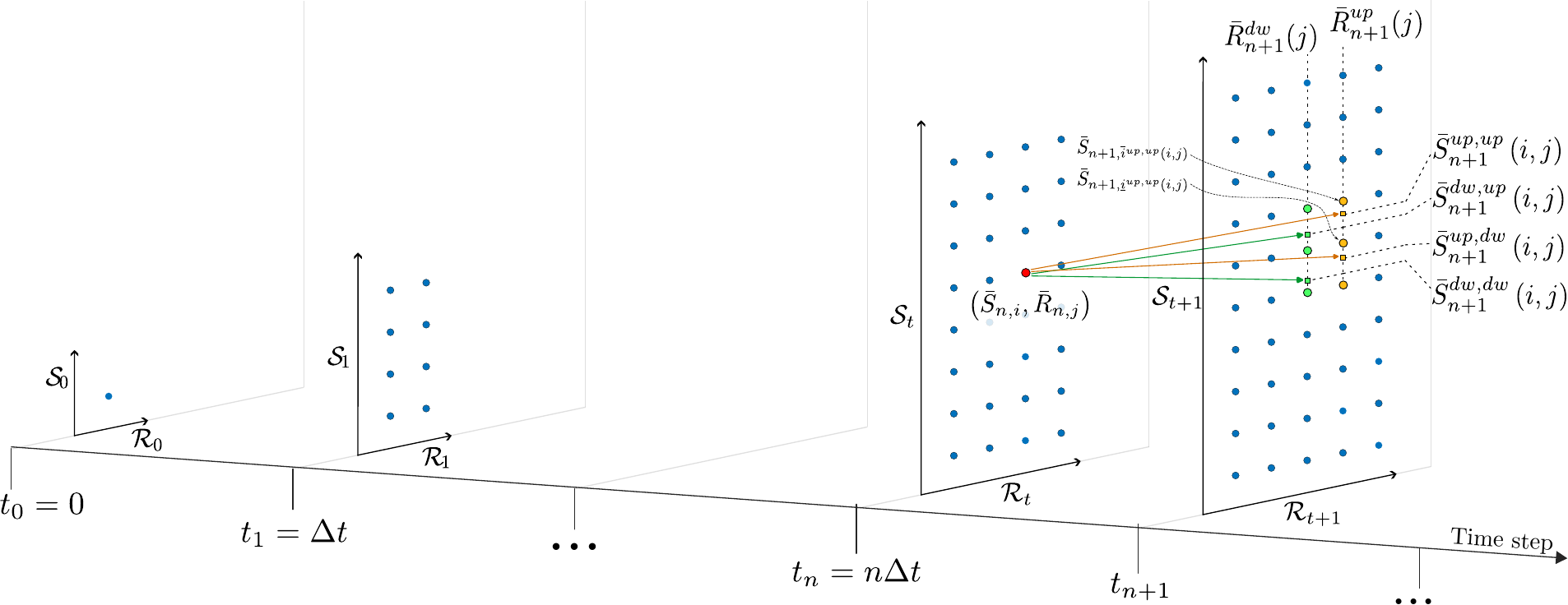}
\par\end{centering}
\caption{\label{fig:TUVHW}Structure of the nodes used by the Tree UVHW algorithm
for evaluating a vanilla option.}
\end{figure}

Figure \ref{fig:TUVHW} summarizes the node structure and the backward
procedure just described. In particular, the figure shows the structure
of the nodes at initial time $t_{0}$, at time $t_{1}$, at a generic
instant $t_{n}$ and at its successor $t_{n+1}$. In addition, for
a generic node $\ensuremath{\left(\bar{R}_{n,j},\bar{S}_{n,i}\right)}$
at time $t_{n}$ (see the red dot), the successor nodes (green and
orange squares) and the grid nodes used for interpolation (green and
orange dots) are highlighted.

On the whole, by applying Equation (\ref{eq:expected}) backward in
time, one can obtain the price of an European option in the Black-Scholes
model with stochastic interest rate described by the Hull-White model. 

Now, recall that in the UVM model, the value of $\sigma$ is defined
at each time instant $t$, for every market state $S_{t},R_{t}$,
in a way that optimizes the contract's continuation value. This is
achieved by applying formula (\ref{eq:expected}), where the function
$f$ represents the value of the contract being evaluated. Specifically,
we solve the following optimization problem 
\begin{align}
\underset{\sigma\in\left[\sigma_{\min},\sigma_{\max}\right]}{\text{min/max}} & \frac{1}{2}p_{n,j}^{\bar{R},up}d_{n}^{up}\left(j\right)\left[f\left(\bar{R}_{n+1}^{up}\left(j\right),\bar{S}_{n+1}^{up,up}\left(i,j\right)\left(\sigma\right)\right)+f\left(\bar{R}_{n+1}^{up}\left(j\right),\bar{S}_{n+1}^{up,dw}\left(i,j\right)\left(\sigma\right)\right)\right]+\nonumber \\
 & \frac{1}{2}p_{n,j}^{\bar{R},dw}d_{n}^{dw}\left(j\right)\left[f\left(\bar{R}_{n+1}^{dw}\left(j\right),\bar{S}_{n+1}^{dw,up}\left(i,j\right)\left(\sigma\right)\right)+f\left(\bar{R}_{n+1}^{dw}\left(j\right),\bar{S}_{n+1}^{dw,dw}\left(i,j\right)\left(\sigma\right)\right)\right],\label{eq:minmax-1}
\end{align}
where the dependence of the future values $\bar{S}_{n+1}^{up,up}\left(i,j\right),\bar{S}_{n+1}^{up,dw}\left(i,j\right),\bar{S}_{n+1}^{dw,up}\left(i,j\right)$
and $\bar{S}_{n+1}^{dw,dw}\left(i,j\right)$ according to equation
(\ref{eq:suu}), (\ref{eq:sud}), (\ref{eq:sdu}) and (\ref{eq:sdd}),
is pointed out by writing $\sigma$ in round brackets. Various numerical
techniques can be employed to tackle this one-dimensional optimization
problem. However, due to the lack of information on the uniqueness
of the maximum, we opted for a straightforward grid search. In this
approach, the objective function is evaluated at $N_{\sigma}$ uniformly
spaced points between $\ensuremath{\sigma_{\min}}$ and $\sigma_{\max}$,
and the best result is selected. Specifically, for $N_{\sigma}=2$,
only the endpoints of the interval $\sigma_{\min}$ and $\sigma_{\max}$
are used. Although this method is simple, it proves effective and
empirical evidence suggests that the optimum frequently occurs at
either $\sigma_{\min}$ or $\sigma_{\max}$. In particular, when $\rho=0$,
it can be shown that the maximum coincides with one of these boundary
values.

In conclusion, by solving problem (\ref{eq:minmax-1}) backward in
time, one can obtain the price of an European option in the UVM with
stochastic interest rate described by the Hull-White model. 
\begin{rem}
The aforementioned procedure for the valuation of European options
can be expanded to encompass the valuation of options with early exercise
feature, such as Bermudian and American options. In particular, at
each time step $t_{n}$, a comparison must be made between the continuation
value and the exercise value of the option, with due care to distinguish
between the possible Short, Long and Mixed approaches, as described
in Section \ref{sec:The-financial-model}. We will describe this procedure
in detail with reference to EIAs in the Section \ref{sec:The-contract}.
\end{rem}

\begin{rem}
As a consequence of Proposition \ref{prop:3}, also $I_{n}=\int_{t_{n}}^{t_{n+1}}R_{s}ds$
is a Gaussian random variable correlated with $R$ and $W^{S}$. Thus,
one could have treated this integral as a Gaussian random variable
to be discretized, as done for $\left(W_{t_{n+1}}^{S}-W_{t}^{S}\mid\bar{R}_{t_{n+1}},\bar{R}_{t}\right)$.
Since $I_{n}$ appears in the definition of the future nodes for $S_{t_{n+1}}$,
according to (\ref{eq:S_u-1}), this discretization would result in
doubling the number of points at which the function $f$ in (\ref{eq:expected})
must be evaluated, thus increasing the computational cost of the algorithm.
However, in the proposed implementation, such an integral has been
discretized using the trapezoidal method. This technique proves to
be more efficient from a numerical point of view. In fact, both the
two ways of handling $I_{n}$ have been implemented and the differences
in the pricing results turns out to be less than $1\permil$. The
discretization of the integral as a Gaussian random variable, on the
other hand, increased the computation time by approximately $30\%$
compared to the discretization using the trapezoidal rule. 
\end{rem}

\section{The EIA contract\label{sec:The-contract}}

Let us discuss the structure of EIAs with cliquet-style guarantees
under specific constraints, including the minimum contract value and
the potential for early surrender. The contract description proposed
in this paper is quite general and accommodates other more specific
formulations, such as those proposed by Kijima and Wong \cite{kijima2007pricing},
Wei et al. \cite{wei2013pricing} and Cui et al. \cite{cui2017equity}.

The contract is structured with periodic monitoring dates and spans
$T$ years, where $T$ is an integer value. The payoff depends on
the returns of the equity index at predetermined uniformly distributed
time points $\left\{ \tilde{t}_{m}=m\frac{T}{M},m=0,\dots,M\right\} $
between $t=0$ and $t=T$. Therefore, the periodic linear return $Y_{m}$
of the underlying index in $\left[\tilde{t}_{m-1},\tilde{t}_{m}\right]$
reads out
\[
Y_{m}=\frac{S_{\tilde{t}_{m}}}{S_{\tilde{t}_{m-1}}}-1,\quad m=1,\ldots,M.
\]
 For each period, the returns are adjusted using a local cap $C_{l}$
and floor $F_{l}$. Additionally, the cumulative return across all
periods is subject to a global cap $C_{g}$ and a floor $F_{g}$.
The resulting value is added to a running total. In this regard, we
define the discrete time processes $Z=\left(Z_{m}\right)_{m\in\left\{ 0,\dots,M\right\} }$
and $X=\left(X_{m}\right)_{m\in\left\{ 0,\dots,M\right\} }$ as follows:
\begin{align}
Z_{m} & =\sum_{j=1}^{m}\max\left(F_{l},\min\left(C_{l},Y_{j}\right)\right),\quad Z_{0}=0,\label{eq:Za}\\
X_{m} & =K\left[H+\max\left(F_{g},\min\left(C_{g},Z_{m}\right)\right)\right],
\end{align}
where $K>0$ is a notional value and $H$ can be either $1$ or $0$
depending on the contract formulation. In particular, the scope of
$H$ is to determine the basis of contract formulation, whether that
be net return or gross return. Please, observe that the cumulative
sum $Z$ reflects the growth of the annuity over time, incorporating
the effects of local caps and floors on the periodic returns. In contrast,
$X$ captures the impact of global caps and floors applied to the
cumulative return. A global minimum contract value (MCV), defined
as
\begin{equation}
G_{m}=\gamma\left[\left(1+g\right)^{\tilde{t}_{m}}+1-H\right],\label{eq:Gi}
\end{equation}
ensures that the payoff the $m-th$ monitoring date cannot fall below
$G_{m}$. More specifically, $g\geq0$ represents the guaranteed minimum
rate, and $\gamma$ is a fraction of the initial premium. In particular,
the payoff at maturity is given by: 
\begin{equation}
\Psi\left(X_{T},G_{T}\right)=\max\left(X_{T},G_{T}\right).\label{eq:payoff}
\end{equation}

Incorporating an early surrender option allows the policyholder to
terminate the contract at specific monitoring dates, if it is considered
financially advantageous. In this case, at monitoring date $m$, the
policyholder can either continue the contract or surrender to receive
a payoff equal to the greater between the accumulated value $Z_{m}$
and the corresponding minimum contract value, that is
\[
\Psi\left(X_{m},G_{m}\right)=\max\left(X_{m},G_{m}\right).
\]
In practice, early surrender often incurs charges that decrease over
time. For example, a surrender penalty $\xi\left(m,N_{c}\right)$
may apply, which usually depends on $N_{c}$, that is the number of
periods the penalty is active. In this particular instance, the payoff
in such cases reads out: 
\begin{equation}
\Psi\left(m,X_{m},G_{m}\right)=\max\left(\left(1-\xi(m,N_{c})\right)X_{m},G_{m}\right).\label{eq:payoff_phi}
\end{equation}

We stress out that the valuation of the EIAs contract with MCV, as
described above, involves pricing a Bermudan path-dependent contingent
claim within a continuous bivariate stochastic model. For this reason,
it is necessary to use a specially designed numerical algorithm, such
as the one described in the next Section.
\begin{rem}
Comparing with the literature on this topic, we can observe that the
contract investigated by Kijima and Wong \cite{kijima2007pricing}
and by Wei et al. \cite{wei2013pricing} can be obtained by setting
$K=1$, $H=1$ and $F_{g}=-\infty$, $C_{g}=+\infty$. Moreover, the
contract discussed by Cui et al. \cite{cui2017equity} can be obtained
by setting $H=0$ and $\gamma=1$.
\end{rem}

\section{The Tree UVHW method for EIAs }

Let us discuss how to adapt the Tree UVHW algorithm to the evaluation
of an EIA contract. First, we recall that, as explained in Section
\ref{sec:The-contract}, the contract's lifespan is divided into $M$
periods at the end of which the value of the underlying equity index
$S$ is observed. For the sake of simplicity, we assume an annual
monitoring frequency, though other frequencies, such as monthly, may
also be considered. Under this assumption, monitoring dates align
with the contract's anniversary, that is $\tilde{t}_{m}=m$, and their
total number $M$ equals $T$. 

We divide each period $\left[m,m+1\right]$ into $N_{L}$ (local)
sub periods, each of duration $\Delta t=1/N_{L}$, so that $N_{T}=N_{L}M$
is the total number of discretization steps from contract inception
to maturity time $T$ and $T=N_{T}\Delta t$. Since $N_{L}$ is an
integer, the monitoring dates $m=1,\dots,M$ are naturally included
within the algorithm time steps. 

The procedure leverages both local and global structures to exploit
the geometric properties of the contract under consideration, thereby
reducing the computational complexity of the problem. Specifically,
we define a structure as \emph{global} if it extends temporally from
the initial time $t=0$ to maturity $t=T$. Conversely, we define
a structure as \emph{local} if it extends temporally between two price
monitoring dates, $m$ and $m+1$.

The first global structure we consider is the tree $\mathcal{R}=\left\{ \mathcal{R}_{n},n=0,\dots,N_{T}\right\} $,
which is employed to discretize the process $R$ through the discrete
time process $\bar{R},$ from $t=0$ and $t=T$ with $N_{T}$ time
steps and time increment $\Delta t$. Such a tree is built as described
in Section \ref{sec:4}.

Then, we define a global structure $\mathcal{Z}=\left\{ \mathcal{Z}_{m},m=0,\dots,M\right\} $,
which is employed to discretize the process $Z$ from $t=0$ and $t=T$
with $M$ time steps and time increment $1$ year. Specifically, we
define a grid $\mathcal{Z}_{m}=\left\{ Z_{m,k}\right\} _{k=1,\dots,N_{Z}}$
of values for the process $Z$ for each monitoring date $m=1,\dots,M$.
Such a grid consists of a set of $N_{Z}$ values uniformly distributed
between $Z_{m,1}=mF_{l}$ and $Z_{m,N_{Z}}=mC_{l}$, that is, the
minimum and maximum possible values for $Z_{m}$, respectively. In
addition, we set $\mathcal{Z}_{0}=\left\{ Z_{0,1}=0\right\} $. 

The third structure we employ is a local grid $\mathcal{S}=\left\{ \mathcal{S}_{h},h=0,\dots,N_{L}\right\} $,
which is used to discretize the process $S$ through the discrete
time process $\bar{S},$ between two monitoring dates $m$ and $m+1$
, with $N_{L}$ time steps per period and time increment $\Delta t$.
This structure $\mathcal{S}$ serves to keep track of the local evolution
of the equity index $S$. In particular, the values $S_{m}$ and $S_{m+1}$
determine the return over the period and thus the new values $Z_{m+1}$
as follows
\[
Z_{m+1}=Z_{m}+\max\left(F_{l},\min\left(C_{l},\alpha\left(\frac{S_{m+1}}{S_{m}}-1\right)\right)\right).
\]
Therefore, knowledge of $Z_{m+1}$ is sufficient to determine the
exercise value of the option at the $\left(m+1\right)-th$ monitoring
date, that is 

\[
\Psi\left(m+1,K\left[H+\max\left(F_{g},\min\left(C_{g},Z_{m+1}\right)\right)\right],G_{m+1}\right).
\]
 Moreover, since the law of the return $\left(S_{m+1}/S_{m}-1\mid S_{m}=s\right)$
is the same of $\left(S_{m+1}-1\mid S_{m}=1\right)$, we can write
\[
Z_{m+1}\sim Z_{m}+\max\left(F_{l},\min\left(C_{l},\alpha\left(S_{m+1}-1\right)\right)\right).
\]
Thus, the process $Z$ is Markovian: it is possible to obtain $Z_{m+1}$
from $Z_{m}$ and $\left(S_{m+1}\mid S_{m}=1\right)$. 

At each monitoring date $m$, that is at time step $mN_{L},$ the
contract value is uniquely determined by the values of processes $Z_{m}$
and $R_{m}$, whereas at an intermediate time $t_{n}=n\Delta t$,
such that $m<t_{n}<m+1$, the contract value is uniquely determined
by the values of processes $Z_{m}$, $R_{t_{n}}$, and $\left(S_{n\Delta t}\mid S_{m}=1\right)$.
Accordingly, for each value $Z_{m,k}$ in $\mathcal{Z}_{m}$, we build
the tree structure $\mathcal{S}$ to discretize the underlying process
between monitoring dates $m$ and $m+1$ under the assumption $S_{m}=1$,
therefore $\mathcal{S}_{0}=\left\{ 1\right\} $. Such a structure
is built as described in Section \ref{sec:4}.

To describe the price of a contract, as approximated by the proposed
method, we use two distinct families of pricing functions: a local
one and a global one. Let us denote with $V_{m}^{glo}$ the approximation
of the contract price at the $m$-th monitoring date (in particular,
the superscript $glo$ stands for \emph{global}). Such a function
depends on the cumulative return process $Z$ and on the discrete
rate process $\bar{R}$: 
\[
V_{m}^{glo}=V_{m}^{glo}\left(Z_{m,k},\bar{R}_{mN_{L},j}\right),\quad Z_{m,k}\in\mathcal{Z}_{m},\bar{R}_{mN_{L},j}\in\mathcal{R}_{mN_{L}}.
\]

We also consider a family of local functions, denoted by $V_{m,h,k}^{loc}$,
determined under the assumption $Z_{m}=Z_{m,k}\in\mathcal{Z}_{m}$,
to represent the contract price between two monitoring dates, say
at time $t_{n}=n\Delta t$ in the time interval $\left[m,m+1\right]$
(in particular, the superscript $loc$ in $V_{m,h,k}^{loc}$ stands
for \emph{local}). In particular, $m=\left\lfloor n/N_{L}\right\rfloor $
is the monitoring date immediately before time $t_{n}=n\Delta t$,
and $h$ is the remainder of the division of $n$ by $N_{L}$, so
that we write $n=mN_{L}+h$. Please, observe that each of these functions
$V_{m,h,k}^{loc}$ is identified by the observation time $m$, the
number of (sub)-steps $h$ after monitoring date $m$, and by $k$
the index of the value $Z_{m,k}\in\mathcal{Z}_{m}$ taken by the process
$Z$ at the $m$-th monitoring date. Each function $V_{m,h,k}^{loc}$
depends on the discrete processes $\bar{R}$ and $\bar{S}$ at the
$n-th$ time step:
\[
V_{m,h,k}^{loc}=V_{m,h,k}^{loc}\left(\bar{R}_{n,j},\bar{S}_{h,i}\right),\quad h\in\left\{ 0,1,\dots,N_{L}\right\} ,\bar{R}_{n,j}\in\mathcal{R}_{n},\bar{S}_{h,i}\in\mathcal{S}_{h}.
\]

Let us now discuss how these functions interact with each other. At
maturity, we observe that the value of $V_{M}^{glo}$ can be deduced
by the payoff function by setting
\[
V_{T}^{glo}\left(Z_{M,k},\bar{R}_{N_{T},j}\right)=\Psi\left(T,K\left[H+\max\left(F_{g},\min\left(C_{g},Z_{M,k}\right)\right)\right],G_{T}\right).
\]
Then, let us assume that the function $V_{m+1}^{glo}$ is known. In
order to compute $V_{m}^{glo}$, we exploit $V_{m,h,k}^{loc}$. Specifically,
for all $k=1,\dots,N_{Z}$, we set 
\begin{equation}
V_{m,N_{L},k}^{loc}\left(\bar{R}_{n,j},\bar{S}_{h,i}\right)=V_{m+1}^{glo}\left(Z_{m,k}+\max\left(F_{l},\min\left(C_{l},\alpha\left(\bar{S}_{h,i}-1\right)\right)\right),\bar{R}_{n,j}\right).\label{eq:glo_loc}
\end{equation}
Please, observe that equation (\ref{eq:glo_loc}) is pivotal in establishing
a connection between the global and local problems. If the value $Z_{m,k}+\max\left(F_{l},\min\left(C_{l},\alpha\left(\bar{S}_{h,i}-1\right)\right)\right)$
is not included among the values of $\mathcal{Z}_{m+1}$, then linear
interpolation of $V_{m+1}^{glo}$ over $\mathcal{Z}_{m+1}\times\mathcal{R}_{\left(m+1\right)N_{T}}$
is used.

Now, let us assume that the function $V_{m,h+1,k}^{loc}$ is known
and let us discuss how to compute $V_{m,h,k}^{loc}$. We focus on
a particular node of the tree, that is $\left(\bar{R}_{n,j},\bar{S}_{h,i}\right)\in\mathcal{R}_{n}\times\mathcal{S}_{h}$.
We compute $V_{m,h,k}^{loc}\left(\bar{R}_{n,j},\bar{S}_{h,i}\right)$
as the solution of the optimization problem (\ref{eq:minmax}) where
$f$ is replaced by $V_{m,h+1,k}^{loc}$, that is
\begin{align}
\underset{\sigma\in\left[\sigma_{\min},\sigma_{\max}\right]}{\text{min/max}} & \frac{1}{2}p_{n,j}^{\bar{R},up}d_{n}^{up}\left(j\right)\left[V_{m,h+1,k}^{loc}\left(\bar{R}_{n+1}^{up}\left(j\right),\bar{S}_{h+1}^{up,up}\left(i,j\right)\left(\sigma\right)\right)+V_{m,h+1,k}^{loc}\left(\bar{R}_{n+1}^{up}\left(j\right),\bar{S}_{h+1}^{up,dw}\left(i,j\right)\left(\sigma\right)\right)\right]+\nonumber \\
 & \frac{1}{2}p_{n,j}^{\bar{R},dw}d_{n}^{dw}\left(j\right)\left[V_{m,h+1,k}^{loc}\left(\bar{R}_{n+1}^{dw}\left(j\right),\bar{S}_{h+1}^{dw,up}\left(i,j\right)\left(\sigma\right)\right)+V_{m,h+1,k}^{loc}\left(\bar{R}_{n+1}^{dw}\left(j\right),\bar{S}_{h+1}^{dw,dw}\left(i,j\right)\left(\sigma\right)\right)\right]\label{eq:minmax}
\end{align}
 where $V_{m,h+1,k}^{loc}$ is extended by linear interpolation over
$\mathcal{R}_{n+1}\times\text{\ensuremath{\mathcal{S}_{h+1}}}$.

We highlight that in (\ref{eq:minmax}), the minimum operator is applied
when analyzing a long position on the contract, while the maximum
operator is used for a short position. Solving problems (\ref{eq:minmax})
iteratively, backward in time, we are able to obtain $V_{m,0,k}^{loc}$
on $\mathcal{R}_{mN_{L}}\times\mathcal{S}_{0}$, where we underline
again $\mathcal{S}_{0}=\left\{ 1\right\} $. Finally, we set 
\begin{equation}
V_{m}^{glo}\left(Z_{m,k},\bar{R}_{mN_{L},j}\right)=V_{m,0,k}^{loc}\left(\bar{R}_{mN_{L},j},1\right),\quad Z_{m,k}\in\mathcal{Z}_{m},\bar{R}_{mN_{L},j}\in\mathcal{R}_{mN_{L}}.\label{EU}
\end{equation}
Please note that Equation (\ref{EU}) acts as a link, thereby enabling
the intermediate solution of the global problem to be obtained from
the final solution of the local problem. This procedure, replicated
iteratively from monitoring date $m=M$ to $m=0$, allows the value
of the European contract to be calculated back in time to the initial
time $t=0$. Thus, the value of the contract at inception is approximated
by $V_{0}^{glo}\left(0,0\right)$.

Furthermore, if the contract permits early surrender, the contract
value at any monitoring date $m$ is defined as the maximum between
the continuation value $\mathcal{C}_{m}$ and the exercise value $\mathcal{E}_{m}$.
In this regard, we replace Equation (\ref{EU}) with the following
one:
\begin{equation}
V_{m}^{glo}\left(Z_{m,k},\bar{R}_{n,j}\right)=\max\left(\mathcal{C}_{m}\left(Z_{m,k},\bar{R}_{n,j}\right),\mathcal{E}_{m}\left(Z_{m,k}\right)\right),\label{eq:AM}
\end{equation}
with 
\begin{align}
\mathcal{C}_{m}\left(Z_{m,k},\bar{R}_{n,j}\right) & =V_{m,0,k}^{loc}\left(\bar{R}_{n,j},1\right)\label{eq:back1}\\
\mathcal{E}_{m}\left(Z_{m,k}\right) & =\Psi\left(m,K\left[H+\max\left(F_{g},\min\left(C_{g},Z_{m,k}\right)\right)\right],G_{m}\right).\label{eq:back2}
\end{align}
Equation (\ref{eq:AM}) holds when we assume a pure long or a pure
short position, that is when we aim to compute $V_{am}^{Long}$ or
$V_{am}^{Short}$ respectively. In contrast, when a mixed position
is considered, we consider operator $\max$ in (\ref{eq:minmax})
and we replace (\ref{eq:AM}) with
\begin{multline}
V_{m}^{glo}\left(Z_{m,k},\bar{R}_{n,j}\right)=\mathcal{C}_{m}^{Short}\left(Z_{m,k},\bar{R}_{n,j}\right)\ind_{\mathcal{C}_{m}^{Long}\left(Z_{m,k},\bar{R}_{n,j}\right)>\mathcal{E}_{m}\left(Z_{m,k}\right)}\\
+\mathcal{E}_{m}\left(Z_{m,k}\right)\ind_{\mathcal{C}_{m}^{Long}\left(Z_{m,k},\bar{R}_{n,j}\right)\leq\mathcal{E}_{m}\left(Z_{m,k}\right)},\label{eq:back3}
\end{multline}
that is, we apply the optimal exercise strategy of the long position
jointly with the contract value maximization provided by the UVM for
the short position.
\begin{rem}
If the contract provides for non-annual monitoring dates, the procedure
described in this Section remains valid, with the only caveat that,
instead of dividing the year into $N_{L}$ sub-periods, the period
between two consecutive monitoring dates $\left[\tilde{t}_{m-1},\tilde{t}_{m}\right]$
will be divided, accordingly.
\end{rem}

\section{The GTU algorithm for EIAs}

To assess the performance of the proposed algorithm, we compare it
with a well-established algorithm. Due to the limited literature in
this field, we have selected another tree-based algorithm, the Gaussian
Process Regression - Tree for the UVM model (GTU), introduced by Goudenège
et al. \cite{goudenege2024leveraging}. Originally developed for the
multidimensional UVM model, this algorithm is particularly effective
in pricing options on multiple underlyings, each modeled by UVM with
uncertain correlations. However, GTU can be adapted to price EIAs
within the stochastic interest rate UVM framework considered in this
study and here we explain how to adapt this algorithm for our purpose.

The UVM algorithm operates backward in time, starting from the contract's
maturity and progressing to the initial time. At each time step, it
computes the contract's price for each point on a random grid by determining
the continuation value as the expected discounted value of the contract
at the subsequent time step, using a tree-based procedure. To extend
the contract's value from the grid to the entire space, the algorithm
employs Gaussian Process Regression (GPR), a machine learning technique
that derives an approximating function from observed values on scattered
points.

Let us now go into the details of the GTU algorithm for pricing EIAs.
As in the previous Section \ref{sec:4}, the valuation algorithm will
be described assuming an annual revaluation basis for the contract.
Using the same notation, we consider a number $N_{L}$ of time steps
per contract period, so that $N_{T}=N_{L}M$ is the total number of
discretization steps. As usual, $\Delta t=1/N_{L}$ is the time interval
between two time steps of the algorithm. The algorithm starts by defining
a stochastic grid of values for each time step, which will subsequently
be utilized for computing the GPR. Therefore, we consider a set of
$N_{MC}$ Monte Carlo paths for the market variables $\left(S,R,I\right)$,
sampled using a time discretization step equal to $\Delta t$, which
we denote with
\[
\left\{ \left(S_{n}^{i},R_{n}^{i},I_{n}^{i}\right),n=0,\dots,N_{T},i=1,\dots,N_{MC}\right\} .
\]
This indicates that $S_{n}^{i}$ represents the specific $i$-th simulated
value of $S_{n\Delta t}$ and the same applies to the other market
variables. Such a simulation can be easily performed by considering
the algorithm described in Goudenège et al. \cite{goudenege2016pricing}
for the Black-Scholes Hull-White model once a particular value for
the volatility $\sigma$ is considered. In particular, at each time
step, we exploit Proposition \ref{prop:3} to simulate the Gaussian
random variables $\left(R_{\left(n+1\right)\Delta t}-R_{n\Delta t}\mid R_{n\Delta t}=R_{n}^{i}\right)$,
$\left(I_{\left(n+1\right)\Delta t}-I_{n\Delta t}\mid R_{n\Delta t}=R_{n}^{i}\right)$
and $W_{\left(n+1\right)\Delta t}^{S}-W_{n\Delta t}^{S}$, and then
we set
\begin{align*}
R_{n+1}^{i} & =\left(R_{\left(n+1\right)\Delta t}-R_{n\Delta t}\right)+R_{n}^{i},\\
S_{n+1}^{i} & =S_{n}^{i}\exp\left(\left(I_{\left(n+1\right)\Delta t}-I_{n\Delta t}\right)-\left(\eta+\frac{\sigma^{2}}{2}\right)\Delta t+\sigma\left(W_{\left(n+1\right)\Delta t}^{S}-W_{n\Delta t}^{S}\right)\right).
\end{align*}

Since in UVM $\sigma$ in constrained in the range $\left[\sigma_{\min},\sigma_{\max}\right]$,
we address the lack of knowledge on such a parameter by splitting
the set of simulations into two equally sized subgroups; half of the
trajectories are simulated using $\sigma=\sigma_{\min}$, while the
other half are simulated using $\sigma=\sigma_{\max}$. After running
these simulations, for each monitoring date $m$ we compute the respective
value for process $Z$, that is 
\[
Z_{m}^{i}=\sum_{j=1}^{m}\max\left(F_{l},\min\left(C_{l},\alpha\left(S_{mN_{L}}^{i}/S_{\left(m-1\right)N_{L}}^{i}-1\right)\right)\right),\quad Z_{0}^{i}=0.
\]
At each time step, we utilize these random simulations to discretize
the market state into a set $\mathcal{X}_{n}$ of specific points.
In particular, let us write $n=mN_{L}+h$ with $m=\left\lfloor n/N_{L}\right\rfloor $
the monitoring date immediately before time $t=n\Delta t$, and $h$
the remainder of the division of $n$ by $N_{L}$. Let us distinguish
between two cases: if $n\Delta t$ is a monitoring date, that is if
$h=0$ and $n=mN_{L}$, then the EIAs price depends only on $Z_{n\Delta t}$
and on $R_{n\Delta t}$. The set designed to illustrate the potential
states of the market is 
\begin{equation}
\mathcal{X}_{m}^{glo}=\left\{ \left(Z_{m}^{i},R_{mN_{L}}^{i}\right),i=1,\dots,N_{MC}\right\} .\label{eq:Xn1}
\end{equation}
On the contrary, if the considered time step is not a monitoring date,
that is if time $t_{n}=n\Delta t$ is between two monitoring dates,
then the EIAs price depends on $Z_{n\Delta t},$ $S_{n\Delta t}/S_{m}$
and on $R_{n\Delta t}$, so we define
\begin{equation}
\mathcal{X}_{n}^{loc}=\left\{ \left(Z_{m},S_{n}^{i}/S_{mN_{L}}^{i},R_{n}^{i}\right),i=1,\dots,N_{MC}\right\} .\label{eq:Xn2}
\end{equation}

The GTU algorithm computes the contract value backward in time, starting
from maturity up to inception. The key elements of this algorithm
are the calculation of continuation values using a tree method and
the use of GPR to ensure that the contract price can also be calculated
from points representing possible market states. Similar to the UVHW
algorithm, we consider both global and local value functions. As in
the previous Section \ref{sec:4}, let us denote $V_{m}^{glo}=V_{m}^{glo}\left(Z_{m}^{i},R_{mN_{L}}^{i}\right)$
the approximation of the contract price at the $m$-th monitoring
date for $m\in\left\{ 0,\dots,M\right\} $. We also consider a family
of functions, denoted by $V_{m,h}^{loc}=V_{m,h}^{loc}\left(Z_{m},S_{n}^{i}/S_{mN_{L}}^{i},R_{n}^{i}\right)$,
that represent the contract price between two monitoring dates, that
is at the $n-th$ time step with $n=mN_{L}+h$ for $m\in\left\{ 0,\dots,M\right\} $
and $h\in\left\{ 0,\dots,N_{L}-1\right\} $.

The contract value at maturity is given by the payoff function $\Psi$,
so we set 
\[
V_{T}^{glo}\left(Z_{M}^{i},R_{N_{T}}^{i}\right)=\Psi\left(T,K\left[H+\max\left(F_{g},\min\left(C_{g},Z_{M}^{i}\right)\right)\right],G_{T}\right).
\]
Then, let us assume that the function $V_{m+1}^{glo}$ is known. In
order to compute $V_{m}^{glo}$, we exploit $V_{m,h}^{loc}$. First
of all let us explain how to compute $V_{m,N_{L}-1}^{loc}$ from $V_{m+1}^{glo}$,
so let us assume $n=\left(m+1\right)N_{L}-1$. The calculation of
$V_{m,N_{L}-1}^{loc}$ for all points in $\mathcal{X}_{n}^{loc}$
requires computing the $\sigma$ value that optimized the continuation
value, that is a conditional expectation. Such an expectation is estimated
by discretizing the possible market states at next time step by exploiting
Proposition \ref{prop:3}. Specifically, let be $C$ the Cholesky
lower factorization of covariance matrix in Proposition \ref{prop:3}.
We discretize the possible Gaussian variations of $R,I$ and $W^{S}$,
that is 
\[
\left(\Delta R_{k}=R_{\left(n+1\right)\Delta t}-R_{n\Delta t}\mid R_{n\Delta t}=R_{n}^{i}\right),\left(\Delta I_{k}=I_{\left(n+1\right)\Delta t}-I_{n\Delta t}\mid R_{n\Delta t}=R_{n}^{i}\right),\left(\Delta W_{k}=W_{\left(n+1\right)\Delta t}^{S}-W_{n\Delta t}^{S}\right)
\]
with eight equiprobable values $\Delta R_{k},\Delta I_{k},\Delta W_{k}$,
$k=1,\dots,8$ defined as
\[
\left(\begin{array}{c}
\Delta R_{k}\\
\Delta I_{k}\\
\Delta W_{k}
\end{array}\right)=\left(\begin{array}{c}
\mu_{R}\\
\mu_{I}\\
\mu_{W}
\end{array}\right)+Cg_{k},
\]
being $g_{k}$ the $k$-th element of the space $\left\{ -1,+1\right\} ^{3}$,
for $k=1,\dots,8$. Thus, for each point in $\mathcal{X}_{n}^{loc}$,
we set
\begin{equation}
V_{m,N_{L}-1}^{loc}\left(Z_{m},S_{n}^{i}/S_{mN_{L}}^{i},R_{n}^{i}\right)=\underset{\sigma\in\left[\sigma_{\min},\sigma_{\max}\right]}{\text{min/max}}\frac{1}{8}\sum_{k=1}^{8}e^{-\Delta I_{k}-\int_{t_{n}}^{t_{n+1}t}\beta\left(s\right)ds}V_{m+1}^{glo}\left(Z_{m+1}^{i,k},R_{n+1}^{i,k}\right)\label{eq:minmaxGPR_1}
\end{equation}
with 
\[
R_{n+1}^{i,k}=R_{n}^{i}+\Delta R_{k}
\]
and
\[
Z_{m+1}^{i,k}=Z_{m}+\max\left(F_{l},\min\left(C_{l},\alpha\left(S_{n}^{i,k}\left(\sigma\right)/S_{mN_{L}}^{i}-1\right)\right)\right),
\]
being
\begin{equation}
S_{n+1}^{i,k}\left(\sigma\right)=S_{n}^{i}\exp\left(\Delta I_{k}+\int_{t_{n}}^{t_{n+1}}\beta\left(s\right)ds-\eta\Delta t-\frac{\sigma^{2}}{2}\Delta t+\sigma\Delta W_{k}\right).\label{eq:S_GPR}
\end{equation}

It is important to observe that in (\ref{eq:minmaxGPR_1}), the points
$\left(Z_{m+1},R_{n+1}^{i,1}\right),\dots,\left(Z_{m+1},R_{n+1}^{i,8}\right)$
at which the function $V_{m+1}^{glo}$ is evaluated do not belong
to $\mathcal{X}_{m+1}^{glo}$ almost sure, so one need to extend $V_{m+1}^{glo}$.
This is done by using the GPR, a powerful Bayesian machine learning
technique used for regression tasks. The reader is referred to Rasmussen
and Williams \cite{rasmussen2006williams} for more information on
GPR. Specifically, the GPR model is trained using $\mathcal{X}_{m+1}^{glo}$
as the predictor set and $\mathcal{Y}_{m+1}^{glo}=\left\{ V_{m+1}^{glo}\left(Z_{m+1}^{i},R_{n+1}^{i}\right),m=1,\dots,M\right\} $
as the target set. Finally, as done for Tree UVHW in Section (\ref{sec:4}),
we solve the optimization problem in (\ref{eq:minmaxGPR_1}) by evaluating
at $N_{\sigma}$ uniformly spaced points between $\ensuremath{\sigma_{\min}}$
and $\sigma_{\max}$, and the best result is selected. 

Now, let us assume that the function $V_{m,h+1}^{loc}$ is known,
and let $n=mN_{L}+h$. The calculation of $V_{m,h}^{glo}$ is developed
following a similar approach to the one devised in (\ref{eq:minmaxGPR_1}).
Specifically, for each point in $\mathcal{X}_{n}^{loc}$, we set

\begin{equation}
V_{m,h}^{glo}\left(Z_{m}^{i},S_{n}^{i}/S_{mN_{L}}^{i},R_{n}^{i}\right)=\underset{\sigma\in\left[\sigma_{\min},\sigma_{\max}\right]}{\text{min/max}}\frac{1}{8}\sum_{k=1}^{8}e^{-\Delta I_{k}-\int_{t_{n}}^{t_{n+1}}\beta\left(s\right)ds}V_{m,h+1}^{loc}\left(Z_{m}^{i},S_{n+1}^{i,k}\left(\sigma\right)/S_{mN_{L}}^{i},R_{n+1}^{i,k}\right),\label{eq:minmaxGPR}
\end{equation}
The points $\left(Z_{m}^{i},S_{n+1}^{i,1}\left(\sigma\right)/S_{mN_{L}}^{i},R_{n+1}^{i,1}\right),\dots,\left(Z_{m},S_{n+1}^{i,8}\left(\sigma\right)/S_{mN_{L}}^{i},R_{n+1}^{i,8}\right)$
at which the function $V_{m,h+1}^{loc}$ is evaluated do not belong
to $\mathcal{X}_{n}^{loc}$ almost sure, so we exploit again GPR.
Specifically, a GPR model is trained using $\mathcal{X}_{n+1}^{loc}$
as the predictor set and $\mathcal{Y}_{n+1}^{loc}=\left\{ V_{m,h+1}^{loc}\left(Z_{m}^{i},S_{n+1}^{i}/S_{mN_{L}}^{i},R_{n+1}^{i}\right),i=1,\dots,N_{MC}\right\} $
as the target set.

Solving problems (\ref{eq:minmaxGPR}) iteratively, backward in time,
we are able to obtain $V_{m,0}^{loc}$ for all points in $\mathcal{X}_{mN_{L}}^{loc}$.
Finally, we define $V_{m}^{glo}$ on $\mathcal{X}_{m}^{glo}$ by setting
\begin{equation}
V_{m}^{glo}\left(Z_{m}^{i},R_{mN_{L}}^{i}\right)=V_{m,0}^{loc}\left(Z_{m}^{i},1,R_{mN_{L}}^{i}\right)\label{EU-1}
\end{equation}
for all points in $\mathcal{X}_{m}^{glo}$.

Finally, it is important to note that the contract under consideration
may include a surrender option. In this case, the relevant price will
be defined by comparing the exercise value with the continuation value,
with appropriate distinctions made between the Long, Short and Mix
formulations discussed in Section \ref{sec:The-financial-model}.
The relevant formulae can be obtained by adapting the corresponding
(\ref{eq:back1}), (\ref{eq:back2}) and (\ref{eq:back3}) developed
for the Tree UVHW algorithm.

\section{Numerical experiments \label{sec:Numerical-experiments}}

This Section presents the results of numerical experiments conducted
to evaluate the performance of the proposed algorithm under various
conditions. In the initial series of tests, the present study evaluates
a contract that is equivalent in parameters to that analyzed in the
recent paper by Cui et al. \cite{cui2017equity}. The objective of
the present study is to emphasize the substantial contribution of
the UVM model and the Hull-White model to the evaluation of contracts
with cliquet-type clauses. In this initial phase of the investigation,
the effects of uncertain volatility and the stochastic rate are considered
separately. In a subsequent phase, however, the joint effect of these
two elements is studied. For the latter, the insurance product treated
by Kijima and Wong \cite{kijima2007pricing} is taken as a point of
reference. The numerical convergence of the Tree UVHW algorithm is
analyzed, after which the valuation of contracts with early surrender
option is studied in detail, with particular attention paid to the
optimal exercise strategy in relation to the model assumptions. 

The evaluation of the EIAs contacts is primarily conducted through
the utilization of the Tree UVHW algorithm. The outcomes of this evaluation
are then compared with those derived via the GTU algorithm. In instances
where feasible, we have additionally presented prices derived through
Monte Carlo simulations, which function as benchmarks. All numerical
procedures were implemented in Matlab, and the calculations were performed
on a PC with 16GB RAM and an Intel Ultra 5 125H processor. 

\subsection{\label{subsec:NT1}Investigating the impact of uncertain volatility
on EIAs price}

\begin{figure}
\begin{centering}
\includegraphics[height=7cm]{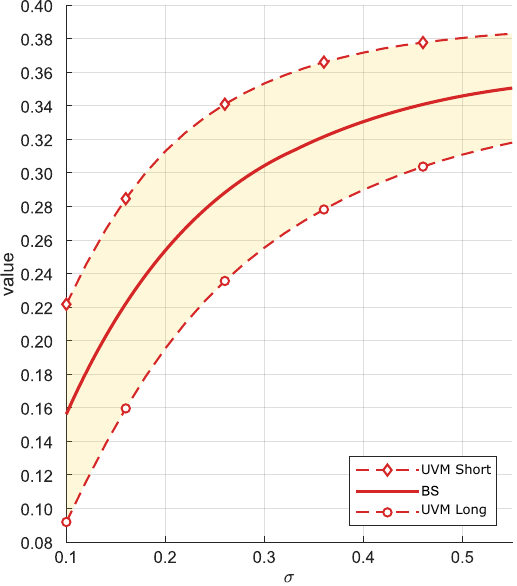}\quad\includegraphics[height=7cm]{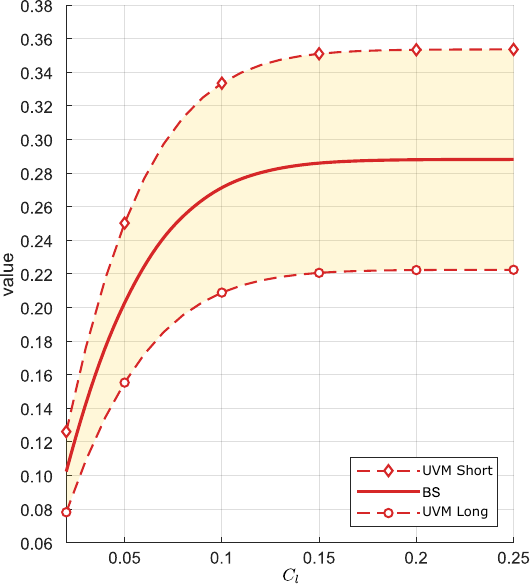}
\par\end{centering}
\caption{\label{fig:F1}Comparison of EIAs prices under BS model and UVM with
constant interest rate. (Left) BS model as $\sigma$ varies and UVM
model with volatility in $\left[\sigma-0.05,\sigma+0.05\right]$ for
$C_{l}=0.08$. (Right) BS model with $\sigma=0.2$ and UVM model with
volatility in $\left[0.15,0.25\right]$.}
\end{figure}

In this sub Section we discuss the disjointed impact of uncertain
volatility under constant interest rate. In this study, the same contract
that was recently examined by Cui et al. \cite{cui2017equity} under
a regime switch model is considered once more. The analysis commences
with a focus on the UVM model, excluding the stochastic rate. This
approach involves the cancellation of rate volatility and the assumption
of a flat curve for ZCB prices. In particular, we compare the result
of the BS model with both minimum and maximum prices returned by the
UVM, to highlight the contribution of uncertain volatility. The parameters
that serve to identify both the contract and the model are enumerated
in Table \ref{tab:parameters-1}. Moreover, the contract provides
for monitoring dates on a monthly rather than annual basis. We stress
out that these parameters align precisely with those used in \cite{cui2017equity}.
It is important to highlight that, for this contract, the returns
are computed on a monthly basis. The resulting graphs are displayed
in Figure \ref{fig:F1}. Specifically, the left panel illustrates
the price as a function of $\sigma$, while the right panel shows
the variation with respect to $C_{l}$. The results are noteworthy
for two key reasons. First, they demonstrate that in UVM, the contract
price is no longer a single fixed value but rather a range of possible
values. The price derived from the BS model falls within this range,
and more broadly, the price is influenced by the evaluator perspective.
Secondly, in contrast to what was observed in Figures 4 and 6 in \cite{cui2017equity}
and remarked by the authors, the price does not collapse into a single
value either for high values of $\sigma$ or for relatively small
values of $C_{l}$. This aspect shows that UVM is capable of revealing
significant differences in prices, which should be taken into account
by risk managers when selling policies. The use of an uncertain volatility
model is able to point out discrepancies in prices that would otherwise
go undetected.

\begin{table}
\begin{centering}
\resizebox{0.9\textwidth}{!}{%
\begin{tabular}{cllccll}
\toprule 
\multicolumn{3}{c}{Model parameters} &  & \multicolumn{3}{c}{Product parameters}\tabularnewline
\cmidrule{1-3} \cmidrule{2-3} \cmidrule{3-3} \cmidrule{5-7} \cmidrule{6-7} \cmidrule{7-7} 
Symbol & Meaning & Value &  & Symbol & Meaning & Value\tabularnewline
\cmidrule{1-3} \cmidrule{2-3} \cmidrule{3-3} \cmidrule{5-7} \cmidrule{6-7} \cmidrule{7-7} 
$S_{0}$ & Initial value of $S$ & $100$ &  & $T$ & Maturity & $1$\tabularnewline
$\sigma$ & Volatility for BS case & variable &  & $M$ & Number of monitoring dates  & $12$\tabularnewline
$\sigma_{\min}$ & Lower bound for $\sigma$ & $\sigma-0.05$ &  & $K$ & Notional value & $1$\tabularnewline
$\sigma_{\max}$ & Upper bound for $\sigma$ & $\sigma+0.05$ &  & $H$ & Net/gross return switch & $0$\tabularnewline
$\kappa$ & Mean rev. speed of i.r. & $1$ &  & $\gamma$ & MCV factor & $1$\tabularnewline
$\omega$ & Volatility of i.r. & $0$ &  & $F_{l}$ & Local floor return & $0$\tabularnewline
$\rho$ & Correlation & $0$ &  & $F_{g}$ & Global floor return & $0$\tabularnewline
$\eta$ & Dividend yield & $0$ &  & $C_{l}$ & Local cap return & variable\tabularnewline
 &  &  &  & $C_{g}$ & Global cap return & $9C_{l}$\tabularnewline
 &  &  &  & $\xi$ & Surrender penalty & $0$\tabularnewline
 &  &  &  & $g$ & guaranteed minimum rate & $3\%$\tabularnewline
 &  &  &  & E. s. & Early surrender & No\tabularnewline
 &  &  &  &  &  & \tabularnewline
\midrule 
\multicolumn{3}{c}{Yield curve parameters} &  & \multicolumn{3}{c}{Algorithm parameters}\tabularnewline
\cmidrule{1-3} \cmidrule{2-3} \cmidrule{3-3} \cmidrule{5-7} \cmidrule{6-7} \cmidrule{7-7} 
Symbol & Meaning & Value &  & Symbol & Meaning & Value\tabularnewline
\cmidrule{1-3} \cmidrule{2-3} \cmidrule{3-3} \cmidrule{5-7} \cmidrule{6-7} \cmidrule{7-7} 
$\beta_{0}$ & Long-term level & $0.05$ &  & $N_{L}$ & N. of time steps per period & $256$\tabularnewline
$\beta_{1}$ & Short-term slope & $0$ &  & $N_{S}$ & N. of grid refinements for $S$ & $16$\tabularnewline
$\beta_{2}$ & Medium-term curvature & $0$ &  & $N_{Z}$ & N. of space points for $Z$ & $16$\tabularnewline
$\beta_{3}$ & Long-term curvature & $0$ &  & $N_{\sigma}$ & N. of values for $\sigma$ optimization & $100$\tabularnewline
$\tau_{1}$ & Short-term decay factor & $1$ &  &  &  & \tabularnewline
$\tau_{2}$ & Long-term decay factor & $1$ &  &  &  & \tabularnewline
\bottomrule
\end{tabular}}\caption{\label{tab:parameters-1} Parameters employed for numerical test in
Sub Section \ref{subsec:NT1}.}
\par\end{centering}
\end{table}

\subsection{\label{subsec:NT2}Investigating the impact of stochastic interest
rate on EIAs price}

\begin{figure}
\begin{centering}
\includegraphics[height=7cm]{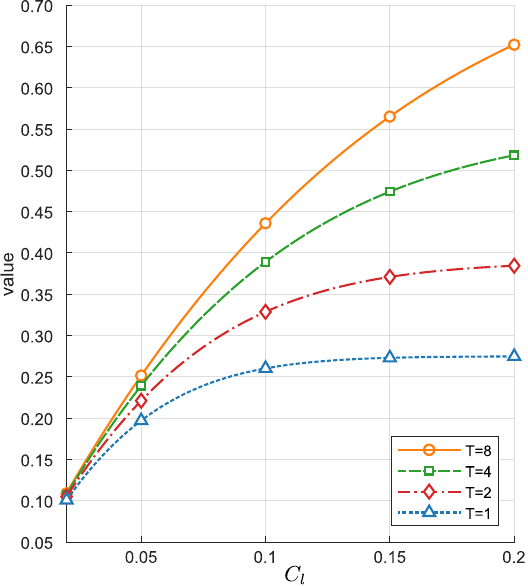}\quad\includegraphics[height=7cm]{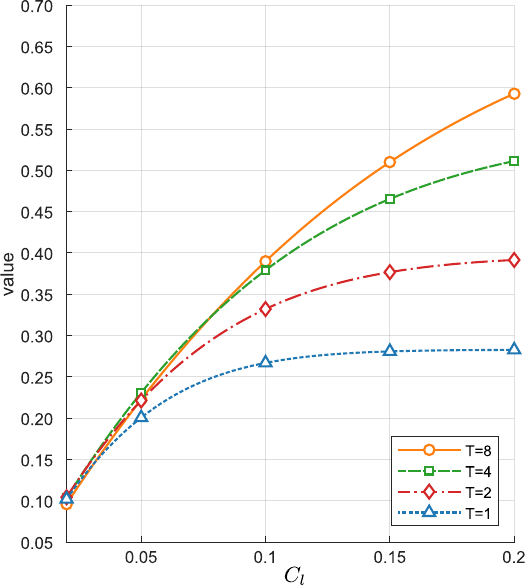}
\par\end{centering}
\caption{\label{fig:F2} Comparison of EIAs prices under BS HW model, under
the yield curve of September 2, 2020 (left) or September 2, 2024 (right).}
\end{figure}

In this sub Section, we examine the impact of a stochastic interest
rate model on the valuation of exchange-traded income annuities under
the assumption of constant volatility. Given that the fund exhibits
constant volatility, we employ the Black-Scholes Hull-White (BS HW)
model. As in the previous sub Section, our study reviews the work
developed by Cui et al. in \cite{cui2017equity}, with a particular
focus on Figure 5 of that paper. The parameters of the algorithm are
given in Table \ref{tab:parameters_2}, while the resulting graphs
are presented in Figure \ref{fig:F2}. Specifically, we examine two
yield curves published by the European Central Bank\footnote{For further details, see \url{https://www.ecb.europa.eu/stats/financial_markets_and_interest_rates/euro_area_yield_curves/html/index.en.html}.}
for the euro area: one from September 2, 2020, representing a low-interest-rate
environment, and another from September 2, 2024, reflecting high-interest-rate
conditions. The resulting graphs share both similarities and differences
with those presented by Cui et al.. Notably, we observe that prices
increase with rising $C_{l}$. In the left graph, contrary to findings
in \cite{cui2017equity}, prices also increase with longer maturities.
A similar trend is observed in the right graph, provided that $\ensuremath{C_{l}}$
is not excessively low.

\begin{table}
\centering{}\resizebox{\textwidth}{!}{%
\begin{tabular}{clcclllllllllrr@{\extracolsep{0pt}.}l}
\toprule 
\multicolumn{2}{c}{Model parameters} &  & \multicolumn{5}{c}{Product parameters} &  & \multicolumn{2}{l}{Algorithm parameters} &  & \multicolumn{4}{c}{Yield curve parameters}\tabularnewline
\cmidrule{1-2} \cmidrule{2-2} \cmidrule{4-8} \cmidrule{5-8} \cmidrule{6-8} \cmidrule{7-8} \cmidrule{8-8} \cmidrule{10-11} \cmidrule{11-11} \cmidrule{13-16} \cmidrule{14-16} \cmidrule{15-16} 
$S_{0}$ & $100$ &  & $T$ & variable &  & $F_{l}$ & $0$ &  & $N_{\tau}$ & $256$ &  & Day & Sep. 2, 2020 & Sep& 2, 2024\tabularnewline
$\sigma$ & $0.20$ &  & $M$ & $12$ &  & $F_{g}$ & $0$ &  & $N_{S}$ & $16$ &  & $\beta_{0}$ & $\phantom{-1}0.269446$ & \multicolumn{2}{c}{$\phantom{-1}0.503770$}\tabularnewline
$\sigma_{\min}$ & $0.20$ &  & $K$ & $1$ &  & $C_{l}$ & variable &  & $N_{Z}$ & $16$ &  & $\beta_{1}$ & $-\phantom{1}0.862654$ & \multicolumn{2}{c}{$\phantom{-1}3.051550$}\tabularnewline
$\sigma_{\max}$ & $0.20$ &  & $H$ & $0$ &  & $C_{g}$ & $10.8C_{l}$ &  & $N_{\sigma}$ & $1$ &  & $\beta_{2}$ & $\phantom{-}12.121820$ & \multicolumn{2}{c}{$-\phantom{1}1.578007$}\tabularnewline
$\kappa$ & $0.2$ &  & $\gamma$ & $1$ &  & E.s. & No &  &  &  &  & $\beta_{3}$ & $-14.133376$ & \multicolumn{2}{c}{$\phantom{-1}6.467092$}\tabularnewline
$\omega$ & $0.03$ &  &  &  &  &  &  &  &  &  &  & $\tau_{1}$ & $\phantom{-1}1.955330$ & \multicolumn{2}{c}{$\phantom{-1}1.710328$}\tabularnewline
$\rho$ & $-0.3$ &  &  &  &  &  &  &  &  &  &  & $\tau_{2}$ & $\phantom{-1}2.105346$ & \multicolumn{2}{c}{$\phantom{-}12.217401$}\tabularnewline
$\eta$ & $0$ &  &  &  &  &  &  &  &  &  &  &  &  & \multicolumn{2}{c}{}\tabularnewline
\bottomrule
\end{tabular}}\caption{\label{tab:parameters_2} Parameters employed for numerical test in
Sub Section \ref{subsec:NT2}. See Table \ref{tab:parameters-1} for
a summary explanation of the symbols listed here.}
\end{table}

\subsection{\label{subsec:NT3}Pricing EIAs in the UVM with stochastic interest
rate without surrender risk}

In this sub Section, we analyze the complete UVM Hull-White model
and apply the Tree UVHW pricing algorithm to evaluate EIA contracts.
We compare our results with those from the milestone paper \cite{kijima2007pricing}
by Kijima and Wong, where the contract is assessed using a stochastic
interest rate model. The contract parameters are provided in Table
\ref{tab:parameters_3}. It is worth noting that, for this contract,
the returns are calculated on an annual basis.

First, we numerically investigate the convergence of the proposed
method by varying both the number of discretization steps in the algorithm
and the correlation coefficient $\rho$. Furthermore, we compute both
the minimum and maximum prices under uncertain volatility, together
with the price under certain volatility as determined by the BS HW
model. The prices obtained using the Tree UVHW algorithm are reported
in Table \ref{tab:C1} for $C_{l}=16\%$ and in Table \ref{tab:C2}
for $C=20\%$. To assess the quality of the results, we compare the
obtained values with those calculated using two benchmark procedures.
The first benchmark is GTU with $M=8000$ random-points and $N_{L}=64$
time steps per period; the second is a classic Monte Carlo with $N_{M}=10^{7}$
simulations, available only in the certain volatility model. In this
latter case, the $95\%$ confidence interval indicates an uncertainty
of less than one hundredth for all the results reported in the Tables.

The results are very positive. In all the considered cases, the prices
returned by the Tree UVHW are very similar to those of the benchmarks,
and moreover, only a few seconds of computation are sufficient to
obtain highly satisfactory results. On the other hand, please note
that the benchmark computed with GTU requires approximately 8 hours
of processing time. Conversely, the Monte Carlo method is very fast:
only a few minutes are needed to obtain results accurate to the hundredth;
however, this method is not applicable in the case of uncertain volatility.
In conclusion, we can state that the proposed method converges quickly
and that the computational time is reasonable, being much lower than
that of GTU. Based on the numerical results just discussed, the algorithm
configuration with $N_{L}=64$, $N_{S}=4$, $N_{Z}=4$, and $N_{\sigma}=2$
returns very accurate results with a reasonable computation time.
We will therefore use it as standard in subsequent tests, unless specified
otherwise.

Now that these prices can be considered correct, let us briefly discuss
the results qualitatively. It can be observed that the price recorded
in the BS HW model invariably lies between the minimum (long) and
maximum (short) prices, as expected. The difference between these
two extreme prices are estimated at $12\%$ and $13\%$, respectively,
for $C=16\%$ and $C=20\%$. Conversely, the impact of the correlation
coefficient appears negligible, with a maximum price change of less
than $1\%$ observed across all cases when the coefficient varies
from $\rho=-0.3$ to $\rho=0.3$.

\begin{table}
\centering{}\resizebox{\textwidth}{!}{%
\begin{tabular}{clcclllllllllr@{\extracolsep{0pt}.}l}
\toprule 
\multicolumn{2}{c}{Model parameters} &  & \multicolumn{5}{c}{Product parameters} &  & \multicolumn{2}{l}{Algorithm parameters} &  & \multicolumn{3}{c}{Yield curve parameters}\tabularnewline
\cmidrule{1-2} \cmidrule{2-2} \cmidrule{4-8} \cmidrule{5-8} \cmidrule{6-8} \cmidrule{7-8} \cmidrule{8-8} \cmidrule{10-11} \cmidrule{11-11} \cmidrule{13-15} \cmidrule{14-15} 
$S_{0}$ & $100$ &  & $T$ & $7$ &  & $F_{l}$ & $0$ &  & $N_{L}$ & variable &  & Day & Sep& 2, 2024\tabularnewline
$\sigma$ & $0.25$ &  & $M$ & $7$ &  & $F_{g}$ & $-\infty$ &  & $N_{S}$ & variable &  & $\beta_{0}$ & \multicolumn{2}{c}{$\phantom{-1}0.503770$}\tabularnewline
$\sigma_{\min}$ & $0.20$ &  & $K$ & $1$ &  & $C_{l}$ & $16\%$ or $20\%$ &  & $N_{Z}$ & variable &  & $\beta_{1}$ & \multicolumn{2}{c}{$\phantom{-1}3.051550$}\tabularnewline
$\sigma_{\max}$ & $0.30$ &  & $H$ & $1$ &  & $C_{g}$ & $+\infty$ &  & $N_{\sigma}$ & variable &  & $\beta_{2}$ & \multicolumn{2}{c}{$-\phantom{1}1.578007$}\tabularnewline
$\kappa$ & $0.2$ &  & $\gamma$ & $0.9$ &  & E.s. & variable &  &  &  &  & $\beta_{3}$ & \multicolumn{2}{c}{$\phantom{-1}6.467092$}\tabularnewline
$\omega$ & $0.03$ &  & $\xi$ & $0$ &  & $g$ & $3\%$ &  &  &  &  & $\tau_{1}$ & \multicolumn{2}{c}{$\phantom{-1}1.710328$}\tabularnewline
$\rho$ & $-0.3,0,+0.3$ &  &  &  &  &  &  &  &  &  &  & $\tau_{2}$ & \multicolumn{2}{c}{$\phantom{-}12.217401$}\tabularnewline
$\eta$ & $0$ &  &  &  &  &  &  &  &  &  &  &  & \multicolumn{2}{c}{}\tabularnewline
\bottomrule
\end{tabular}}\caption{\label{tab:parameters_3} Parameters employed for numerical test in
Sub Section \ref{subsec:NT3}. See Table \ref{tab:parameters-1} for
a summary explanation of the symbols listed here.}
\end{table}

\begin{table}[p]
\begin{centering}
\renewcommand{\arraystretch}{0.7}%
\begin{tabular}{llcccccccccccc}
\toprule 
 &  &  & \multicolumn{3}{c}{Min price (Long)} &  & \multicolumn{3}{c}{BS HW price} &  & \multicolumn{3}{c}{Max price (Short)}\tabularnewline
 &  & $\rho$ & $-0.3$ & $0$ & $0.3$ &  & $-0.3$ & $0$ & $0.3$ &  & $-0.3$ & $0$ & $0.3$\tabularnewline
\midrule
 &  &  &  &  &  &  &  &  &  &  &  &  & \tabularnewline
\multicolumn{14}{l}{Tree method with $N_{S}=4$, $N_{Z}=4$, $N_{\sigma}=2$}\tabularnewline
\midrule
$N_{L}$ & $16$ &  & {\small{}$\underset{\left(1\right)}{\pt{1.1466}}$} & {\small{}$\underset{\left(1\right)}{\pt{1.1404}}$} & {\small{}$\underset{\left(1\right)}{\pt{1.1398}}$} &  & {\small{}$\underset{\left(1\right)}{\pt{1.2107}}$} & {\small{}$\underset{\left(1\right)}{\pt{1.2048}}$} & {\small{}$\underset{\left(1\right)}{\pt{1.2012}}$} &  & {\small{}$\underset{\left(1\right)}{\pt{1.2772}}$} & {\small{}$\underset{\left(1\right)}{\pt{1.2742}}$} & {\small{}$\underset{\left(1\right)}{\pt{1.2659}}$}\tabularnewline
 & $32$ &  & {\small{}$\underset{\left(5\right)}{\pt{1.1447}}$} & {\small{}$\underset{\left(5\right)}{\pt{1.1403}}$} & {\small{}$\underset{\left(5\right)}{\pt{1.1379}}$} &  & {\small{}$\underset{\left(3\right)}{\pt{1.2098}}$} & {\small{}$\underset{\left(3\right)}{\pt{1.2048}}$} & {\small{}$\underset{\left(3\right)}{\pt{1.2004}}$} &  & {\small{}$\underset{\left(5\right)}{\pt{1.2775}}$} & {\small{}$\underset{\left(5\right)}{\pt{1.2728}}$} & {\small{}$\underset{\left(5\right)}{\pt{1.2663}}$}\tabularnewline
 & $64$ &  & {\small{}$\underset{\left(28\right)}{\pt{1.1439}}$} & {\small{}$\underset{\left(30\right)}{\pt{1.1397}}$} & {\small{}$\underset{\left(28\right)}{\pt{1.1371}}$} &  & {\small{}$\underset{\left(18\right)}{\pt{1.2097}}$} & {\small{}$\underset{\left(17\right)}{\pt{1.2049}}$} & {\small{}$\underset{\left(17\right)}{\pt{1.2003}}$} &  & {\small{}$\underset{\left(29\right)}{\pt{1.2782}}$} & {\small{}$\underset{\left(30\right)}{\pt{1.2731}}$} & {\small{}$\underset{\left(29\right)}{\pt{1.2670}}$}\tabularnewline
 & $128$ &  & {\small{}$\underset{\left(189\right)}{\pt{1.1435}}$} & {\small{}$\underset{\left(190\right)}{\pt{1.1393}}$} & {\small{}$\underset{\left(204\right)}{\pt{1.1367}}$} &  & {\small{}$\underset{\left(284\right)}{\pt{1.2095}}$} & {\small{}$\underset{\left(310\right)}{\pt{1.2047}}$} & {\small{}$\underset{\left(236\right)}{\pt{1.2001}}$} &  & {\small{}$\underset{\left(313\right)}{\pt{1.2784}}$} & {\small{}$\underset{\left(498\right)}{\pt{1.2731}}$} & {\small{}$\underset{\left(461\right)}{\pt{1.2672}}$}\tabularnewline
\multicolumn{14}{l}{Tree method with $N_{S}=8$, $N_{Z}=8$, $N_{\sigma}=8$}\tabularnewline
\midrule
$N_{L}$ & $16$ &  & {\small{}$\underset{\left(24\right)}{\pt{1.1463}}$} & {\small{}$\underset{\left(7\right)}{\pt{1.1410}}$} & {\small{}$\underset{\left(23\right)}{\pt{1.1396}}$} &  & {\small{}$\underset{\left(3\right)}{\pt{1.2108}}$} & {\small{}$\underset{\left(2\right)}{\pt{1.2046}}$} & {\small{}$\underset{\left(2\right)}{\pt{1.2013}}$} &  & {\small{}$\underset{\left(19\right)}{\pt{1.2775}}$} & {\small{}$\underset{\left(7\right)}{\pt{1.2739}}$} & {\small{}$\underset{\left(21\right)}{\pt{1.2660}}$}\tabularnewline
 & $32$ &  & {\small{}$\underset{\left(128\right)}{\pt{1.1445}}$} & {\small{}$\underset{\left(34\right)}{\pt{1.1405}}$} & {\small{}$\underset{\left(122\right)}{\pt{1.1378}}$} &  & {\small{}$\underset{\left(25\right)}{\pt{1.2101}}$} & {\small{}$\underset{\left(26\right)}{\pt{1.2050}}$} & {\small{}$\underset{\left(25\right)}{\pt{1.2007}}$} &  & {\small{}$\underset{\left(149\right)}{\pt{1.2781}}$} & {\small{}$\underset{\left(33\right)}{\pt{1.2730}}$} & {\small{}$\underset{\left(74\right)}{\pt{1.2668}}$}\tabularnewline
 & $64$ &  & {\small{}$\underset{\left(459\right)}{\pt{1.1435}}$} & {\small{}$\underset{\left(105\right)}{\pt{1.1398}}$} & {\small{}$\underset{\left(405\right)}{\pt{1.1368}}$} &  & {\small{}$\underset{\left(65\right)}{\pt{1.2098}}$} & {\small{}$\underset{\left(65\right)}{\pt{1.2049}}$} & {\small{}$\underset{\left(66\right)}{\pt{1.2003}}$} &  & {\small{}$\underset{\left(404\right)}{\pt{1.2785}}$} & {\small{}$\underset{\left(104\right)}{\pt{1.2731}}$} & {\small{}$\underset{\left(403\right)}{\pt{1.2672}}$}\tabularnewline
 & $128$ &  & {\small{}$\underset{\left(3532\right)}{\pt{1.1430}}$} & {\small{}$\underset{\left(785\right)}{\pt{1.1394}}$} & {\small{}$\underset{\left(3158\right)}{\pt{1.1363}}$} &  & {\small{}$\underset{\left(488\right)}{\pt{1.2096}}$} & {\small{}$\underset{\left(492\right)}{\pt{1.2048}}$} & {\small{}$\underset{\left(657\right)}{\pt{1.2002}}$} &  & {\small{}$\underset{\left(4310\right)}{\pt{1.2787}}$} & {\small{}$\underset{\left(2094\right)}{\pt{1.2732}}$} & {\small{}$\underset{\left(4157\right)}{\pt{1.2675}}$}\tabularnewline
\multicolumn{14}{l}{Benchmarks}\tabularnewline
\midrule
MC &  &  & {\small{}$ $} & {\small{}$ $} & {\small{}$ $} &  & {\small{}$\pt{1.2094}$} & {\small{}$\pt{1.2046}$} & {\small{}$\pt{1.2000}$} &  & {\small{}$ $} & {\small{}$ $} & {\small{}$ $}\tabularnewline
GTU &  &  & {\small{}$\pt{1.1437}$} & {\small{}$\pt{1.1402}$} & {\small{}$\pt{1.1369}$} &  & {\small{}$\pt{1.2103}$} & {\small{}$\pt{1.2056}$} & {\small{}$\pt{1.2010}$} &  & {\small{}$\pt{1.2796}$} & {\small{}$\pt{1.2739}$} & {\small{}$\pt{1.2684}$}\tabularnewline
\bottomrule
\end{tabular}
\par\end{centering}
\caption{\label{tab:C1}Convergence of the Tree UVHW method with local cap
$C_{l}=16\%$. Early surrender is not allowed. The values in parentheses
represent the computational times, measured in seconds.}
\end{table}

\begin{table}[p]
\begin{centering}
\renewcommand{\arraystretch}{0.7}%
\begin{tabular}{llcccccccccccc}
\toprule 
 &  &  & \multicolumn{3}{c}{Min price (Long)} &  & \multicolumn{3}{c}{BS-HW price} &  & \multicolumn{3}{c}{Max price (Short)}\tabularnewline
 &  & $\rho$ & $-0.3$ & $0$ & $0.3$ &  & $-0.3$ & $0$ & $0.3$ &  & $-0.3$ & $0$ & $0.3$\tabularnewline
\midrule
 &  &  &  &  &  &  &  &  &  &  &  &  & \tabularnewline
\multicolumn{14}{l}{Tree method with $N_{S}=4$, $N_{Z}=4$, $N_{\sigma}=2$}\tabularnewline
\midrule
$N_{L}$ & $16$ &  & {\small{}$\underset{\left(1\right)}{\pt{1.1943}}$} & {\small{}$\underset{\left(1\right)}{\pt{1.1892}}$} & {\small{}$\underset{\left(1\right)}{\pt{1.1864}}$} &  & {\small{}$\underset{\left(1\right)}{\pt{1.2693}}$} & {\small{}$\underset{\left(1\right)}{\pt{1.2614}}$} & {\small{}$\underset{\left(1\right)}{\pt{1.2584}}$} &  & {\small{}$\underset{\left(1\right)}{\pt{1.3467}}$} & {\small{}$\underset{\left(1\right)}{\pt{1.3402}}$} & {\small{}$\underset{\left(1\right)}{\pt{1.3338}}$}\tabularnewline
 & $32$ &  & {\small{}$\underset{\left(5\right)}{\pt{1.1926}}$} & {\small{}$\underset{\left(5\right)}{\pt{1.1868}}$} & {\small{}$\underset{\left(5\right)}{\pt{1.1847}}$} &  & {\small{}$\underset{\left(3\right)}{\pt{1.2686}}$} & {\small{}$\underset{\left(3\right)}{\pt{1.2632}}$} & {\small{}$\underset{\left(3\right)}{\pt{1.2578}}$} &  & {\small{}$\underset{\left(5\right)}{\pt{1.3471}}$} & {\small{}$\underset{\left(5\right)}{\pt{1.3422}}$} & {\small{}$\underset{\left(5\right)}{\pt{1.3342}}$}\tabularnewline
 & $64$ &  & {\small{}$\underset{\left(29\right)}{\pt{1.1917}}$} & {\small{}$\underset{\left(29\right)}{\pt{1.1865}}$} & {\small{}$\underset{\left(30\right)}{\pt{1.1838}}$} &  & {\small{}$\underset{\left(18\right)}{\pt{1.2682}}$} & {\small{}$\underset{\left(18\right)}{\pt{1.2628}}$} & {\small{}$\underset{\left(18\right)}{\pt{1.2574}}$} &  & {\small{}$\underset{\left(30\right)}{\pt{1.3473}}$} & {\small{}$\underset{\left(30\right)}{\pt{1.3417}}$} & {\small{}$\underset{\left(30\right)}{\pt{1.3346}}$}\tabularnewline
 & $128$ &  & {\small{}$\underset{\left(197\right)}{\pt{1.1912}}$} & {\small{}$\underset{\left(210\right)}{\pt{1.1863}}$} & {\small{}$\underset{\left(344\right)}{\pt{1.1834}}$} &  & {\small{}$\underset{\left(128\right)}{\pt{1.2680}}$} & {\small{}$\underset{\left(129\right)}{\pt{1.2626}}$} & {\small{}$\underset{\left(129\right)}{\pt{1.2573}}$} &  & {\small{}$\underset{\left(218\right)}{\pt{1.3475}}$} & {\small{}$\underset{\left(219\right)}{\pt{1.3415}}$} & {\small{}$\underset{\left(214\right)}{\pt{1.3348}}$}\tabularnewline
\multicolumn{14}{l}{Tree method with $N_{S}=8$, $N_{Z}=8$, $N_{\sigma}=8$}\tabularnewline
\midrule
$N_{L}$ & $16$ &  & {\small{}$\underset{\left(10\right)}{\pt{1.1943}}$} & {\small{}$\underset{\left(3\right)}{\pt{1.1897}}$} & {\small{}$\underset{\left(10\right)}{\pt{1.1865}}$} &  & {\small{}$\underset{\left(2\right)}{\pt{1.2697}}$} & {\small{}$\underset{\left(2\right)}{\pt{1.2612}}$} & {\small{}$\underset{\left(2\right)}{\pt{1.2589}}$} &  & {\small{}$\underset{\left(11\right)}{\pt{1.3472}}$} & {\small{}$\underset{\left(3\right)}{\pt{1.3406}}$} & {\small{}$\underset{\left(11\right)}{\pt{1.3341}}$}\tabularnewline
 & $32$ &  & {\small{}$\underset{\left(64\right)}{\pt{1.1923}}$} & {\small{}$\underset{\left(18\right)}{\pt{1.1872}}$} & {\small{}$\underset{\left(64\right)}{\pt{1.1845}}$} &  & {\small{}$\underset{\left(11\right)}{\pt{1.2688}}$} & {\small{}$\underset{\left(11\right)}{\pt{1.2634}}$} & {\small{}$\underset{\left(11\right)}{\pt{1.2580}}$} &  & {\small{}$\underset{\left(63\right)}{\pt{1.3474}}$} & {\small{}$\underset{\left(17\right)}{\pt{1.3424}}$} & {\small{}$\underset{\left(66\right)}{\pt{1.3346}}$}\tabularnewline
 & $64$ &  & {\small{}$\underset{\left(422\right)}{\pt{1.1912}}$} & {\small{}$\underset{\left(108\right)}{\pt{1.1867}}$} & {\small{}$\underset{\left(425\right)}{\pt{1.1834}}$} &  & {\small{}$\underset{\left(68\right)}{\pt{1.2684}}$} & {\small{}$\underset{\left(68\right)}{\pt{1.2629}}$} & {\small{}$\underset{\left(68\right)}{\pt{1.2576}}$} &  & {\small{}$\underset{\left(419\right)}{\pt{1.3477}}$} & {\small{}$\underset{\left(111\right)}{\pt{1.3419}}$} & {\small{}$\underset{\left(420\right)}{\pt{1.3349}}$}\tabularnewline
 & $128$ &  & {\small{}$\underset{\left(3175\right)}{\pt{1.1907}}$} & {\small{}$\underset{\left(792\right)}{\pt{1.1864}}$} & {\small{}$\underset{\left(3010\right)}{\pt{1.1829}}$} &  & {\small{}$\underset{\left(490\right)}{\pt{1.2681}}$} & {\small{}$\underset{\left(500\right)}{\pt{1.2627}}$} & {\small{}$\underset{\left(498\right)}{\pt{1.2574}}$} &  & {\small{}$\underset{\left(3030\right)}{\pt{1.3478}}$} & {\small{}$\underset{\left(798\right)}{\pt{1.3416}}$} & {\small{}$\underset{\left(3038\right)}{\pt{1.3351}}$}\tabularnewline
\multicolumn{14}{l}{Benchmarks}\tabularnewline
\midrule
MC &  &  & {\small{}$ $} & {\small{}$ $} & {\small{}$ $} &  & {\small{}$\pt{1.2679}$} & {\small{}$\pt{1.2624}$} & {\small{}$\pt{1.2572}$} &  & {\small{}$ $} & {\small{}$ $} & {\small{}$ $}\tabularnewline
GTU &  &  & {\small{}$\pt{1.1915}$} & {\small{}$\pt{1.1875}$} & {\small{}$\pt{1.1836}$} &  & {\small{}$\pt{1.2691}$} & {\small{}$\pt{1.2636}$} & {\small{}$\pt{1.2583}$} &  & {\small{}$\pt{1.3488}$} & {\small{}$\pt{1.3422}$} & {\small{}$\pt{1.3360}$}\tabularnewline
\bottomrule
\end{tabular}
\par\end{centering}
\caption{\label{tab:C2}Convergence of the Tree UVHW method with local cap
$C_{l}=20\%$. Early surrender is not allowed. The values in parentheses
represent the computational times, measured in seconds.}
\end{table}
\FloatBarrier

\subsection{Pricing EIAs in the UVM with stochastic interest rate with surrender
risk}

We proceed by investigating the convergence of the Tree UVHW in scenarios
where early surrender is a possibility. The outcomes of this analysis
are presented in Tables \ref{tab:C3} and \ref{tab:C4}, respectively,
for $C_{l}=16\%$ and $C_{l}=20\%$. A comparative analysis is then
conducted with the results obtained from GTU and Monte Carlo methods
(this latter only in the case of constant volatility), utilizing an
optimization approach based on the Longstaff-Schwartz method. The
prices returned by the Tree UVHW are found to be in close agreement
with those obtained from the benchmark methods, and good approximations
of the limiting values can be achieved within a matter of seconds.
For the sake of completeness, it should be noted that the computational
times required to compute the benchmark prices are very high, amounting
to several hours for both GTU and Monte Carlo.

Beyond numerical considerations, it is interesting to observe that
the final price is significantly affected by the chosen strategy:
moving from the long strategy to the short strategy increases the
price by approximately $10\%$ for $C_{l}=16\%$ and $12\%$ for $C_{l}=20\%$.
Moreover, we can observe that the short strategy invariably returns
a higher price than the mixed strategy, as anticipated, though the
disparity is marginal, amounting to less than $1\%$ for both values
of $C_{l}$. Moreover, in a manner similar to the context of prohibited
early surrender, the influence of the correlation coefficient $\rho$
appears to be negligible. Finally, it should be noted that, as is
natural, all prices obtained by allowing early surrender are higher
than the respective prices without such an option, reported in Tables
\ref{tab:C1} and \ref{tab:C2}. In particular, the increase varies
between $1\%$ and $2\%$. This also applies when comparing the mixed
strategy with the short strategy without early exercise.
\begin{center}
\begin{table}[p]
\begin{centering}
\resizebox{\textwidth}{!}{\renewcommand{\arraystretch}{0.8}%
\begin{tabular}{llcccccccccccccccc}
\toprule 
 &  &  & \multicolumn{3}{c}{Min price (Long)} &  & \multicolumn{3}{c}{BS HW price} &  & \multicolumn{3}{c}{Mixed strategy} &  & \multicolumn{3}{c}{Max price (Short)}\tabularnewline
 &  & $\rho$ & $-0.3$ & $0$ & $0.3$ &  & $-0.3$ & $0$ & $0.3$ &  & $-0.3$ & $0$ & $0.3$ &  & $-0.3$ & $0$ & $0.3$\tabularnewline
\midrule
 &  &  &  &  &  &  &  &  &  &  &  &  &  &  &  &  & \tabularnewline
\multicolumn{14}{l}{Tree method with $N_{S}=4$, $N_{Z}=4$, $N_{\sigma}=2$} &  &  &  & \tabularnewline
\midrule
$N_{L}$ & $16$ &  & $\underset{\left(1\right)}{\pt{1.1719}}$ & $\underset{\left(1\right)}{\pt{1.1692}}$ & $\underset{\left(1\right)}{\pt{1.1716}}$ &  & $\underset{\left(1\right)}{\pt{1.2292}}$ & $\underset{\left(1\right)}{\pt{1.2261}}$ & $\underset{\left(1\right)}{\pt{1.2258}}$ &  & $\underset{\left(2\right)}{\pt{1.2833}}$ & $\underset{\left(2\right)}{\pt{1.2821}}$ & $\underset{\left(2\right)}{\pt{1.2771}}$ &  & $\underset{\left(1\right)}{\pt{1.2911}}$ & $\underset{\left(1\right)}{\pt{1.2904}}$ & $\underset{\left(1\right)}{\pt{1.2850}}$\tabularnewline
 & $32$ &  & $\underset{\left(5\right)}{\pt{1.1701}}$ & $\underset{\left(5\right)}{\pt{1.1692}}$ & $\underset{\left(5\right)}{\pt{1.1699}}$ &  & $\underset{\left(3\right)}{\pt{1.2283}}$ & $\underset{\left(3\right)}{\pt{1.2263}}$ & $\underset{\left(3\right)}{\pt{1.2250}}$ &  & $\underset{\left(10\right)}{\pt{1.2832}}$ & $\underset{\left(10\right)}{\pt{1.2809}}$ & $\underset{\left(10\right)}{\pt{1.2774}}$ &  & $\underset{\left(5\right)}{\pt{1.2912}}$ & $\underset{\left(5\right)}{\pt{1.2890}}$ & $\underset{\left(5\right)}{\pt{1.2853}}$\tabularnewline
 & $64$ &  & $\underset{\left(30\right)}{\pt{1.1693}}$ & $\underset{\left(30\right)}{\pt{1.1686}}$ & $\underset{\left(30\right)}{\pt{1.1692}}$ &  & $\underset{\left(18\right)}{\pt{1.2282}}$ & $\underset{\left(18\right)}{\pt{1.2264}}$ & $\underset{\left(18\right)}{\pt{1.2249}}$ &  & $\underset{\left(60\right)}{\pt{1.2836}}$ & $\underset{\left(60\right)}{\pt{1.2810}}$ & $\underset{\left(60\right)}{\pt{1.2779}}$ &  & $\underset{\left(31\right)}{\pt{1.2919}}$ & $\underset{\left(31\right)}{\pt{1.2893}}$ & $\underset{\left(31\right)}{\pt{1.2860}}$\tabularnewline
 & $128$ &  & $\underset{\left(199\right)}{\pt{1.1689}}$ & $\underset{\left(198\right)}{\pt{1.1683}}$ & $\underset{\left(197\right)}{\pt{1.1688}}$ &  & $\underset{\left(119\right)}{\pt{1.2280}}$ & $\underset{\left(119\right)}{\pt{1.2263}}$ & $\underset{\left(115\right)}{\pt{1.2247}}$ &  & $\underset{\left(392\right)}{\pt{1.2838}}$ & $\underset{\left(389\right)}{\pt{1.2810}}$ & $\underset{\left(387\right)}{\pt{1.2779}}$ &  & $\underset{\left(196\right)}{\pt{1.2920}}$ & $\underset{\left(197\right)}{\pt{1.2893}}$ & $\underset{\left(198\right)}{\pt{1.2861}}$\tabularnewline
\multicolumn{14}{l}{Tree method with $N_{S}=8$, $N_{Z}=8$, $N_{\sigma}=8$} &  &  &  & \tabularnewline
\midrule
$N_{L}$ & $16$ &  & $\underset{\left(10\right)}{\pt{1.1717}}$ & $\underset{\left(3\right)}{\pt{1.1698}}$ & $\underset{\left(9\right)}{\pt{1.1716}}$ &  & $\underset{\left(2\right)}{\pt{1.2293}}$ & $\underset{\left(2\right)}{\pt{1.2258}}$ & $\underset{\left(2\right)}{\pt{1.2259}}$ &  & $\underset{\left(19\right)}{\pt{1.2834}}$ & $\underset{\left(6\right)}{\pt{1.2817}}$ & $\underset{\left(19\right)}{\pt{1.2771}}$ &  & $\underset{\left(10\right)}{\pt{1.2912}}$ & $\underset{\left(3\right)}{\pt{1.2902}}$ & $\underset{\left(10\right)}{\pt{1.2851}}$\tabularnewline
 & $32$ &  & $\underset{\left(56\right)}{\pt{1.1700}}$ & $\underset{\left(16\right)}{\pt{1.1694}}$ & $\underset{\left(60\right)}{\pt{1.1699}}$ &  & $\underset{\left(10\right)}{\pt{1.2286}}$ & $\underset{\left(10\right)}{\pt{1.2264}}$ & $\underset{\left(10\right)}{\pt{1.2252}}$ &  & $\underset{\left(114\right)}{\pt{1.2834}}$ & $\underset{\left(31\right)}{\pt{1.2808}}$ & $\underset{\left(115\right)}{\pt{1.2778}}$ &  & $\underset{\left(58\right)}{\pt{1.2917}}$ & $\underset{\left(16\right)}{\pt{1.2892}}$ & $\underset{\left(55\right)}{\pt{1.2857}}$\tabularnewline
 & $64$ &  & $\underset{\left(375\right)}{\pt{1.1691}}$ & $\underset{\left(100\right)}{\pt{1.1687}}$ & $\underset{\left(379\right)}{\pt{1.1690}}$ &  & $\underset{\left(59\right)}{\pt{1.2282}}$ & $\underset{\left(60\right)}{\pt{1.2264}}$ & $\underset{\left(61\right)}{\pt{1.2249}}$ &  & $\underset{\left(745\right)}{\pt{1.2836}}$ & $\underset{\left(197\right)}{\pt{1.2809}}$ & $\underset{\left(747\right)}{\pt{1.2779}}$ &  & $\underset{\left(381\right)}{\pt{1.2920}}$ & $\underset{\left(100\right)}{\pt{1.2893}}$ & $\underset{\left(377\right)}{\pt{1.2861}}$\tabularnewline
 & $128$ &  & $\underset{\left(3056\right)}{\pt{1.1686}}$ & $\underset{\left(803\right)}{\pt{1.1682}}$ & $\underset{\left(3054\right)}{\pt{1.1685}}$ &  & $\underset{\left(498\right)}{\pt{1.2281}}$ & $\underset{\left(503\right)}{\pt{1.2263}}$ & $\underset{\left(501\right)}{\pt{1.2248}}$ &  & $\underset{\left(6282\right)}{\pt{1.2835}}$ & $\underset{\left(1592\right)}{\pt{1.2809}}$ & $\underset{\left(7520\right)}{\pt{1.2780}}$ &  & $\underset{\left(3153\right)}{\pt{1.2922}}$ & $\underset{\left(802\right)}{\pt{1.2893}}$ & $\underset{\left(3060\right)}{\pt{1.2863}}$\tabularnewline
\multicolumn{14}{l}{Benchmarks} &  &  &  & \tabularnewline
\midrule
MC &  &  & $ $ & $ $ & $ $ &  & $\pt{1.2268}$ & $\pt{1.2248}$ & $\pt{1.2231}$ &  & $ $ & $ $ & $ $ &  & $ $ & $ $ & $ $\tabularnewline
GTU &  &  & $\pt{1.1693}$ & $\pt{1.1690}$ & $\pt{1.1691}$ &  & $\pt{1.2288}$ & $\pt{1.2271}$ & $\pt{1.2256}$ &  & $\pt{1.2845}$ & $\pt{1.2816}$ & $\pt{1.2788}$ &  & $\pt{1.2931}$ & $\pt{1.2901}$ & $\pt{1.2872}$\tabularnewline
\bottomrule
\end{tabular}}
\par\end{centering}
\caption{\label{tab:C3}Convergence of the Tree UVHW method with local cap
$C_{l}=16\%$. Early surrender is allowed at monitoring dates. The
values in parentheses represent the computational times, measured
in seconds.}
\end{table}
 
\par\end{center}

\begin{center}
\begin{table}[p]
\begin{centering}
\resizebox{\textwidth}{!}{\renewcommand{\arraystretch}{0.8}%
\begin{tabular}{llcccccccccccccccc}
\toprule 
 &  &  & \multicolumn{3}{c}{Min price (Long)} &  & \multicolumn{3}{c}{BS HW price} &  & \multicolumn{3}{c}{Mixed strategy} &  & \multicolumn{3}{c}{Max price (Short)}\tabularnewline
 &  & $\rho$ & $-0.3$ & $0$ & $0.3$ &  & $-0.3$ & $0$ & $0.3$ &  & $-0.3$ & $0$ & $0.3$ &  & $-0.3$ & $0$ & $0.3$\tabularnewline
\midrule
 &  &  &  &  &  &  &  &  &  &  &  &  &  &  &  &  & \tabularnewline
\multicolumn{14}{l}{Tree method with $N_{S}=4$, $N_{Z}=4$, $N_{\sigma}=2$} &  &  &  & \tabularnewline
\midrule
$N_{L}$ & $16$ &  & $\underset{\left(1\right)}{\pt{1.2128}}$ & $\underset{\left(1\right)}{\pt{1.2112}}$ & $\underset{\left(1\right)}{\pt{1.2118}}$ &  & $\underset{\left(1\right)}{\pt{1.2824}}$ & $\underset{\left(1\right)}{\pt{1.2772}}$ & $\underset{\left(1\right)}{\pt{1.2778}}$ &  & $\underset{\left(2\right)}{\pt{1.3480}}$ & $\underset{\left(2\right)}{\pt{1.3438}}$ & $\underset{\left(3\right)}{\pt{1.3414}}$ &  & $\underset{\left(1\right)}{\pt{1.3563}}$ & $\underset{\left(1\right)}{\pt{1.3523}}$ & $\underset{\left(1\right)}{\pt{1.3487}}$\tabularnewline
 & $32$ &  & $\underset{\left(5\right)}{\pt{1.2111}}$ & $\underset{\left(5\right)}{\pt{1.2090}}$ & $\underset{\left(5\right)}{\pt{1.2102}}$ &  & $\underset{\left(3\right)}{\pt{1.2817}}$ & $\underset{\left(3\right)}{\pt{1.2794}}$ & $\underset{\left(3\right)}{\pt{1.2772}}$ &  & $\underset{\left(10\right)}{\pt{1.3483}}$ & $\underset{\left(9\right)}{\pt{1.3456}}$ & $\underset{\left(10\right)}{\pt{1.3413}}$ &  & $\underset{\left(5\right)}{\pt{1.3566}}$ & $\underset{\left(5\right)}{\pt{1.3543}}$ & $\underset{\left(5\right)}{\pt{1.3491}}$\tabularnewline
 & $64$ &  & $\underset{\left(30\right)}{\pt{1.2103}}$ & $\underset{\left(30\right)}{\pt{1.2087}}$ & $\underset{\left(30\right)}{\pt{1.2093}}$ &  & $\underset{\left(18\right)}{\pt{1.2813}}$ & $\underset{\left(18\right)}{\pt{1.2790}}$ & $\underset{\left(18\right)}{\pt{1.2768}}$ &  & $\underset{\left(59\right)}{\pt{1.3485}}$ & $\underset{\left(59\right)}{\pt{1.3452}}$ & $\underset{\left(60\right)}{\pt{1.3414}}$ &  & $\underset{\left(31\right)}{\pt{1.3568}}$ & $\underset{\left(31\right)}{\pt{1.3538}}$ & $\underset{\left(30\right)}{\pt{1.3494}}$\tabularnewline
 & $128$ &  & $\underset{\left(205\right)}{\pt{1.2099}}$ & $\underset{\left(203\right)}{\pt{1.2085}}$ & $\underset{\left(199\right)}{\pt{1.2089}}$ &  & $\underset{\left(118\right)}{\pt{1.2811}}$ & $\underset{\left(117\right)}{\pt{1.2788}}$ & $\underset{\left(120\right)}{\pt{1.2767}}$ &  & $\underset{\left(389\right)}{\pt{1.3486}}$ & $\underset{\left(387\right)}{\pt{1.3449}}$ & $\underset{\left(391\right)}{\pt{1.3412}}$ &  & $\underset{\left(194\right)}{\pt{1.3569}}$ & $\underset{\left(197\right)}{\pt{1.3536}}$ & $\underset{\left(198\right)}{\pt{1.3496}}$\tabularnewline
\multicolumn{14}{l}{Tree method with $N_{S}=8$, $N_{Z}=8$, $N_{\sigma}=8$} &  &  &  & \tabularnewline
\midrule
$N_{L}$ & $16$ &  & $\underset{\left(9\right)}{\pt{1.2129}}$ & $\underset{\left(3\right)}{\pt{1.2117}}$ & $\underset{\left(9\right)}{\pt{1.2120}}$ &  & $\underset{\left(2\right)}{\pt{1.2828}}$ & $\underset{\left(2\right)}{\pt{1.2769}}$ & $\underset{\left(2\right)}{\pt{1.2782}}$ &  & $\underset{\left(20\right)}{\pt{1.3487}}$ & $\underset{\left(6\right)}{\pt{1.3445}}$ & $\underset{\left(20\right)}{\pt{1.3410}}$ &  & $\underset{\left(10\right)}{\pt{1.3567}}$ & $\underset{\left(3\right)}{\pt{1.3527}}$ & $\underset{\left(10\right)}{\pt{1.3490}}$\tabularnewline
 & $32$ &  & $\underset{\left(66\right)}{\pt{1.2110}}$ & $\underset{\left(17\right)}{\pt{1.2094}}$ & $\underset{\left(64\right)}{\pt{1.2101}}$ &  & $\underset{\left(11\right)}{\pt{1.2819}}$ & $\underset{\left(11\right)}{\pt{1.2796}}$ & $\underset{\left(10\right)}{\pt{1.2774}}$ &  & $\underset{\left(114\right)}{\pt{1.3484}}$ & $\underset{\left(31\right)}{\pt{1.3458}}$ & $\underset{\left(116\right)}{\pt{1.3412}}$ &  & $\underset{\left(59\right)}{\pt{1.3568}}$ & $\underset{\left(17\right)}{\pt{1.3544}}$ & $\underset{\left(60\right)}{\pt{1.3494}}$\tabularnewline
 & $64$ &  & $\underset{\left(379\right)}{\pt{1.2100}}$ & $\underset{\left(99\right)}{\pt{1.2089}}$ & $\underset{\left(382\right)}{\pt{1.2091}}$ &  & $\underset{\left(61\right)}{\pt{1.2814}}$ & $\underset{\left(59\right)}{\pt{1.2791}}$ & $\underset{\left(58\right)}{\pt{1.2770}}$ &  & $\underset{\left(772\right)}{\pt{1.3485}}$ & $\underset{\left(200\right)}{\pt{1.3452}}$ & $\underset{\left(1032\right)}{\pt{1.3414}}$ &  & $\underset{\left(423\right)}{\pt{1.3571}}$ & $\underset{\left(108\right)}{\pt{1.3539}}$ & $\underset{\left(416\right)}{\pt{1.3497}}$\tabularnewline
 & $128$ &  & $\underset{\left(3004\right)}{\pt{1.2095}}$ & $\underset{\left(796\right)}{\pt{1.2086}}$ & $\underset{\left(3042\right)}{\pt{1.2087}}$ &  & $\underset{\left(494\right)}{\pt{1.2812}}$ & $\underset{\left(501\right)}{\pt{1.2788}}$ & $\underset{\left(494\right)}{\pt{1.2768}}$ &  & $\underset{\left(6054\right)}{\pt{1.3484}}$ & $\underset{\left(1585\right)}{\pt{1.3449}}$ & $\underset{\left(6067\right)}{\pt{1.3412}}$ &  & $\underset{\left(3037\right)}{\pt{1.3571}}$ & $\underset{\left(798\right)}{\pt{1.3536}}$ & $\underset{\left(3035\right)}{\pt{1.3498}}$\tabularnewline
\multicolumn{14}{l}{Benchmarks} &  &  &  & \tabularnewline
\midrule
MC &  &  & $ $ & $ $ & $ $ &  & $\pt{1.2805}$ & $\pt{1.2780}$ & $\pt{1.2756}$ &  & $ $ & $ $ & $ $ &  & $ $ & $ $ & $ $\tabularnewline
GTU &  &  & $\pt{1.2103}$ & $\pt{1.2096}$ & $\pt{1.2094}$ &  & $\pt{1.2821}$ & $\pt{1.2797}$ & $\pt{1.2776}$ &  & $\pt{1.3496}$ & $\pt{1.3457}$ & $\pt{1.3423}$ &  & $\pt{1.3581}$ & $\pt{1.3543}$ & $\pt{1.3507}$\tabularnewline
\bottomrule
\end{tabular}}
\par\end{centering}
\caption{\label{tab:C4}Convergence of the Tree UVHW method with local cap
$C_{l}=20\%$. Early surrender is allowed at monitoring dates. The
values in parentheses represent the computational times, measured
in seconds.}
\end{table}
 
\par\end{center}

We conclude this sub Section by analyzing the optimal operating region
for a contract with $C_{l}=16\%$ and $\rho=-0.3$. In particular,
we study the qualitative properties that emerge when comparing the
optimal exercise strategies under both long and short positions. It
is interesting to note that, given the parameters in Table \ref{tab:parameters-1},
the exercise value at monitoring date $m$ for this contract is given
by 
\[
\Psi\left(X_{m},G_{m}\right)=\max\left(1+Z_{m},\left(1+g\right)^{m}\right).
\]
 Therefore, the optimal strategy is to exercise the option at the
first contract anniversary $m$ such that the continuation value $\mathcal{C}_{m}\left(Z_{m},R_{m}\right)$
is less than $1+Z_{m}$, $\left(1+g\right)^{m}$, or both. For this
reason, in the graphs shown in Figure \ref{fig:OSR}, three cases
are highlighted: in the red region the inequality $\left(1+g\right)^{m}\geq\mathcal{C}_{m}\left(Z_{m},R_{m}\right)>1+Z_{m}$
is satisfied, in the blue region $1+Z_{m}\geq\mathcal{C}_{m}\left(Z_{m},R_{m}\right)>\left(1+g\right)^{m}$
and in the green region both inequalities $\left(1+g\right)^{m}\geq\mathcal{C}_{m}\left(Z_{m},R_{m}\right)$
and $1+Z_{m}\geq\mathcal{C}_{m}\left(Z_{m},R_{m}\right)$ are satisfied.
Finally, the white region corresponds to points where early exercise
is not convenient, that is $\mathcal{C}_{m}\left(Z_{m},R_{m}\right)>\Psi\left(X_{m},G_{m}\right)$.
In particular, these graphs illustrate the optimal choices as a function
of the value of $Z_{m}\in\left[0,mC_{l}\right]$ and the interest
rate $r_{m}\in\left[-2\%,12\%\right]$ (used in place of $R_{m}$
to facilitate the economic interpretation of the results). We also
distinguish the optimal strategy for a long position (left) from that
for a short position (right).

In general, it can be stated that, for all cases considered, it is
advantageous to terminate the contract early when the interest rate
$r$ is high. The reason for this phenomenon lies in the local cap
on growth, $C_{l}$, which limits the contract's value appreciation
despite high growth rates of the underlying equity index under risk-neutral
probability. This effect is more pronounced in the case of a short
position: the policyholder, fearing a loss in the contract's value,
prefers to exit early, favoring certainty over uncertainty. 

The dependence on $Z_{m}$ is less evident: as $Z_{m}$ increases,
both the continuation value and the linear factor $1+Z_{m}$ grow,
while the constant term $\ensuremath{\left(1+g\right)^{m}}$ remains
unaffected by $Z_{m}$. Moreover, the growth of the continuation value
is sublinear because the payoff is capped due to the cap on $Z_{m}$.
Consequently, for high values of $Z_{m}$, early exercise is advantageous
due to the $1+Z_{m}$ factor (observe the blue region), while, for
low values of $Z_{m}$, it is beneficial due to the $\left(1+g\right)^{m}$
factor (observe the red region). For intermediate values of $Z_{m}$,
early exercise may be advantageous due to both factors (observe the
green region) or disadvantageous (observe the white region), depending
on the specific circumstances.

Finally, it is important to note that the convenience of early exercise
is limited in the initial years, while it becomes more feasible as
time progresses. This is due to the fact that by engaging in exercise
at an early stage, an individual relinquishes the future benefits
provided by the floor in all subsequent monitoring dates.
\begin{center}
\begin{figure}
\begin{centering}
\subfloat[time $t=6$ years.]{\centering{}\includegraphics[scale=0.33]{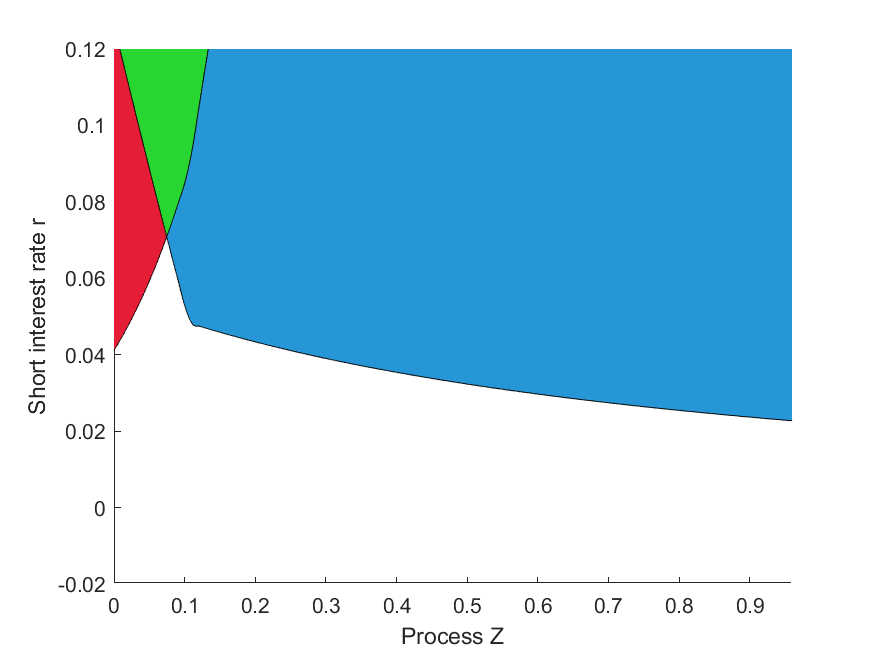}\includegraphics[scale=0.33]{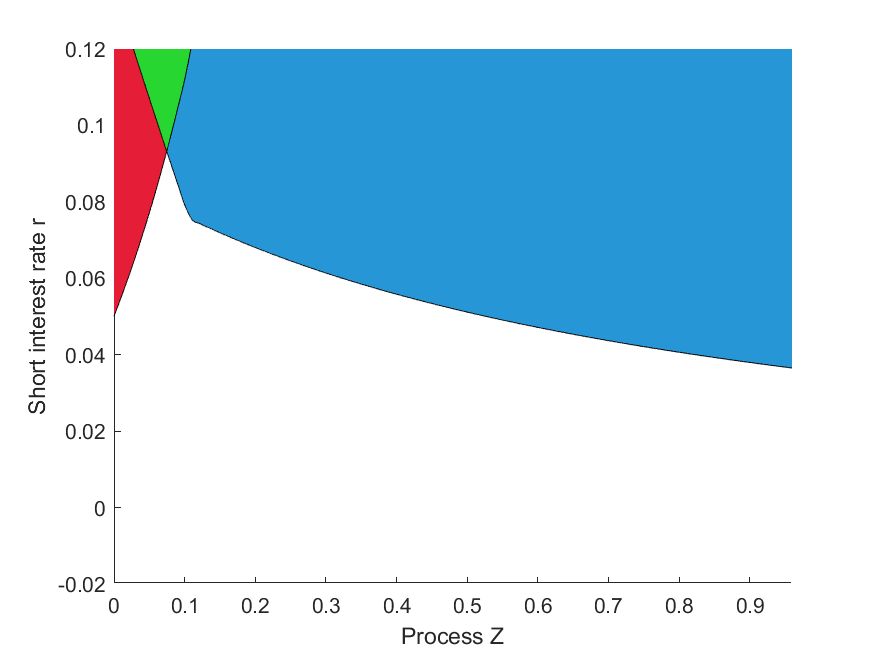}}
\par\end{centering}
\begin{centering}
\subfloat[time $t=4$ years.]{\centering{}\includegraphics[scale=0.33]{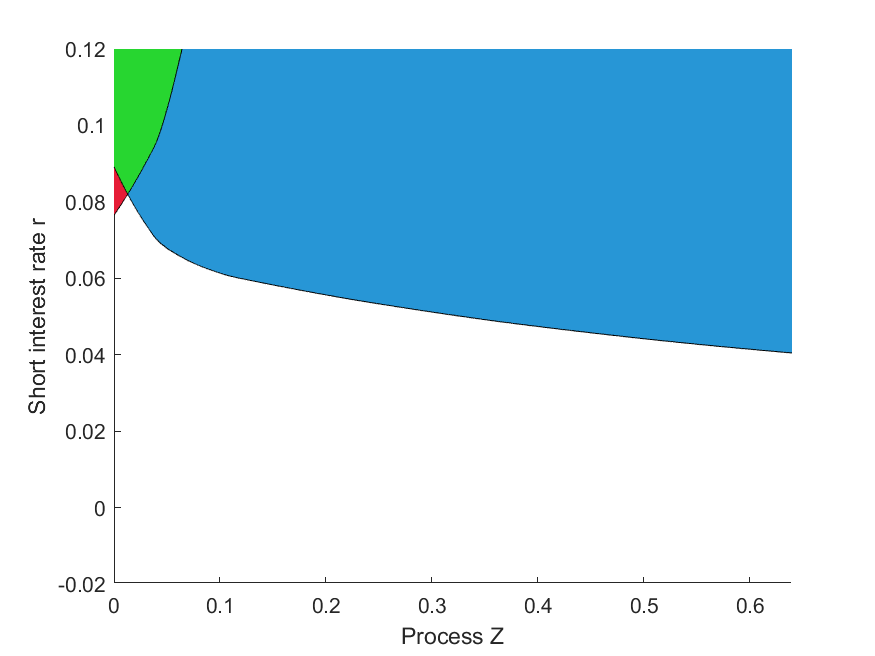}\includegraphics[scale=0.33]{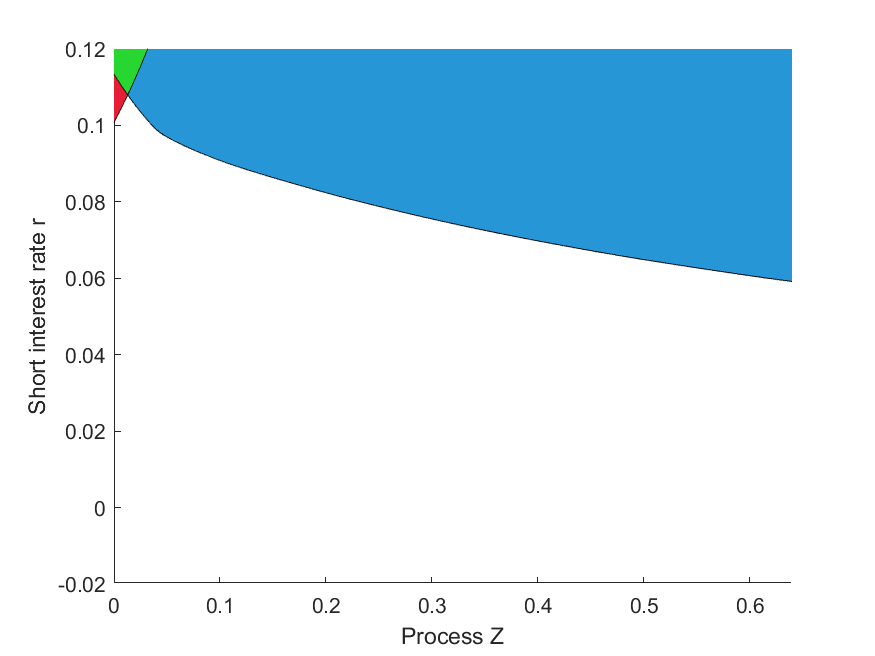}}
\par\end{centering}
\begin{centering}
\subfloat[time $t=2$ years.]{\centering{}\includegraphics[scale=0.33]{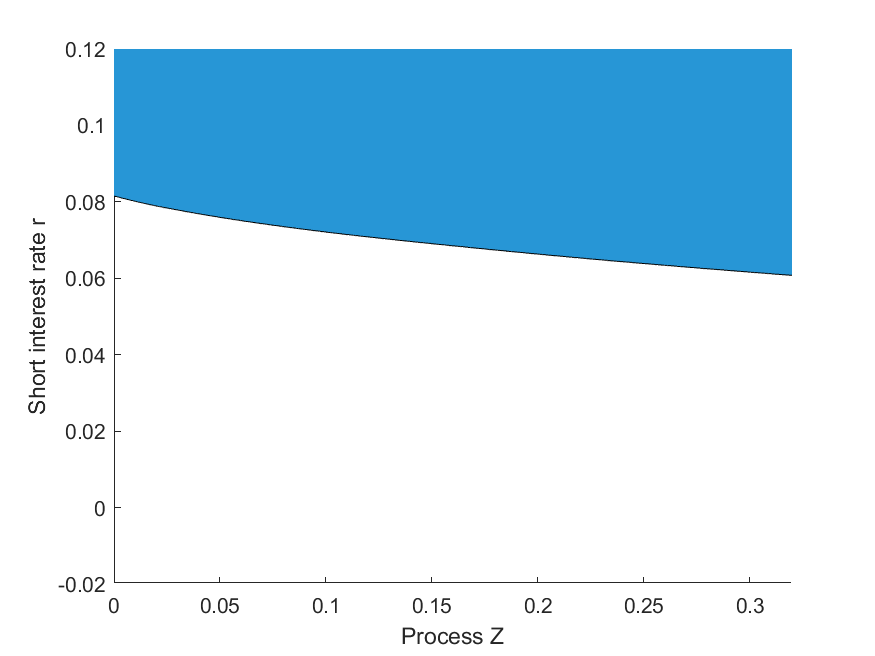}\includegraphics[scale=0.33]{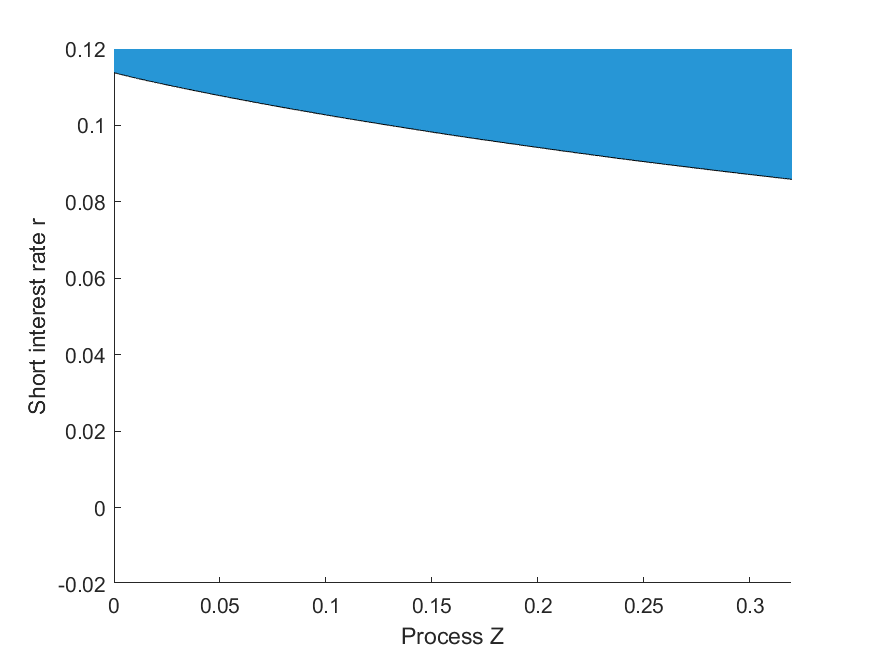}}
\par\end{centering}
\caption{\label{fig:OSR}Optimal surrender region for long (left) and short
(right) position for different monitoring dates. The colored region
denotes where early exercise is convenient. }
\end{figure}
 
\par\end{center}

\FloatBarrier

\section{Conclusion}

In this paper, we have developed a novel tree-based numerical method
for pricing equity-indexed annuities, featuring cliquet-style guarantees
and early surrender risk. Our model integrates an uncertain volatility
framework for the underlying asset with a stochastic interest rate
modeled by the Hull-White dynamics, thereby providing a robust approach
for addressing market uncertainties. This aspect highlights that UVM
enables robust pricing by uncovering significant price differences,
which risk managers should consider when selling policies. Additionally,
we have shown that it is important to consider a stochastic interest
rate model capable of accurately capturing rate dynamics, which play
a crucial role in these long-maturity insurance products. The proposed
Tree UVHW algorithm has been extensively tested and demonstrates high
accuracy and computational efficiency when compared to benchmark methods
such as Monte Carlo simulations and the GTU algorithm. Our numerical
experiments confirm that both volatility uncertainty and the stochastic
behavior of interest rates play a significant role in determining
the value of these complex financial products. In particular, the
inclusion of an early surrender option leads to differentiated optimal
exercise strategies, with distinct implications for short and long
positions. Overall, our findings underscore the importance of employing
flexible modeling frameworks and efficient numerical techniques in
the valuation of exotic, path-dependent contracts.

\appendix

\section{\label{sec:Appendix_A}Appendix: proof of Proposition \ref{prop:3}}

Ostrovski \cite{ostrovski2013efficient} has already demonstrated
that the random variables $\left(R_{t+\Delta t}-R_{t}\mid R_{t}\right)$
and $\left(I_{t+\Delta t}-I_{t}\mid R_{t}\right)$ follow a Gaussian
distribution, as well as established their associated mean, variance,
and covariance indicators. In addition, by definition, $W_{t+\Delta t}^{S}-W_{t}^{S}$
follows a Gaussian distribution, with zero mean and variance equal
to $\Delta t$, which is independent by $R_{t}$, so it has the same
law of $\left(W_{t+\Delta t}^{S}-W_{t}^{S}\mid R_{t}\right)$. Only
the covariance formulas between $\left(W_{t+\Delta t}^{S}-W_{t}^{S}\mid R_{t}\right)$
and the other two random variables remain to be proved. We start by
computing the correlations between the increment $\left(W_{t+\Delta t}^{r}-W_{t}^{r}\mid R_{t}\right)$
and the other two random variables.

The process $R$ follows the Vasicek model, and standard calculations
for this type of process (see e.g. Brigo and Mercurio \cite{brigo2001interest})
allow solving the associated SDE, yielding:
\[
R_{t+\Delta t}=e^{-\kappa\Delta t}\left(R_{t}+\int_{t}^{t+\Delta t}e^{\kappa\left(s-t\right)}dW_{s}^{r}\right).
\]
Therefore, we can write
\begin{align*}
\mathbb{C}\mathrm{ov}\left[\left(R_{t+\Delta t}-R_{t}\mid R_{t}\right),\left(W_{t+\Delta t}^{r}-W_{t}^{r}\mid R_{t}\right)\right] & =\mathbb{C}\mathrm{ov}\left[e^{-\kappa\Delta t}\int_{t}^{t+\Delta t}e^{\kappa\left(s-t\right)}dW_{s}^{r},\int_{t}^{t+\Delta t}dW_{s}^{r}\right]\\
 & =e^{-\kappa\Delta t}\mathbb{C}\mathrm{ov}\left[\int_{t}^{t+\Delta t}e^{\kappa\left(s-t\right)}dW_{s}^{r},\int_{t}^{t+\Delta t}dW_{s}^{r}\right]\\
 & =e^{-\kappa\Delta t}\int_{t}^{t+\Delta t}e^{\kappa\left(s-t\right)}ds\\
 & =\frac{1-e^{-\kappa\Delta t}}{\kappa}.
\end{align*}

\begin{align*}
\mathbb{C}\mathrm{ov}\left[\left(I_{t+\Delta t}-I_{t}\mid R_{t}\right),\left(W_{t+\Delta t}^{r}-W_{t}^{r}\mid R_{t}\right)\right] & =\mathbb{C}\mathrm{ov}\left[\int_{t}^{t+\Delta t}R_{s}ds,\int_{t}^{t+\Delta t}dW_{s}^{r}\right]\\
 & =\mathbb{C}\mathrm{ov}\left[\int_{t}^{t+\Delta t}e^{-\kappa\left(s-t\right)}\left(R_{t}+\int_{t}^{s}e^{\kappa\left(u-t\right)}dW_{u}^{r}\right)ds,\int_{t}^{t+\Delta t}dW_{s}^{r}\right]\\
 & =\mathbb{C}\mathrm{ov}\left[\int_{t}^{t+\Delta t}\int_{t}^{s}e^{-\kappa\left(s-u\right)}dW_{u}^{r}ds,\int_{t}^{t+\Delta t}dW_{s}^{r}\right]\\
 & =\mathbb{C}\mathrm{ov}\left[\int_{t}^{t+\Delta t}\int_{u}^{t+\Delta t}e^{-\kappa\left(s-u\right)}dsdW_{u}^{r},\int_{t}^{t+\Delta t}dW_{s}^{r}\right]\\
 & =\mathbb{C}\mathrm{ov}\left[\frac{1}{\kappa}\int_{t}^{t+\Delta t}1-e^{-\kappa\left(t+\Delta t-u\right)}dW_{u}^{r},\int_{t}^{t+\Delta t}dW_{s}^{r}\right]\\
 & =\frac{1}{\kappa}\int_{t}^{t+\Delta t}1-e^{-\kappa\left(t+\Delta t-u\right)}du\\
 & =\frac{1}{\kappa^{2}}\left(\kappa\Delta t+e^{-\kappa\Delta t}-1\right)
\end{align*}

We emphasize that the interchange of the two integrals, which occurs
in the fourth line of the previous calculation, is justified by the
Fubini-Tonelli Theorem. Specifically, this theorem allows the reversal
of the order of integration in double integrals under certain conditions.
In this case, the function $e^{-\kappa\left(s-u\right)}$ is continuous
and integrable over the region of integration, which is the triangular
domain defined by $t\leq u\leq s\leq t+\Delta t$. This region is
measurable, and the integrand is non-negative and bounded. Therefore,
we can safely exchange the order of integration without violating
any conditions for the application of Fubini's Theorem. This leads
to the simplified expression for the covariance calculation.

Now, we come back to the Gaussian increment $W_{t+\Delta t}^{S}-W_{t}^{s}$.
Since $W^{S}$ and $W^{r}$ are correlated, we can write 
\[
W_{t+\Delta t}^{S}-W_{t}^{S}=\rho\left(W_{t+\Delta t}^{r}-W_{t}^{r}\right)+\sqrt{1-\rho^{2}}G
\]
with $G\sim\mathcal{N}\left(0,\Delta t\right)$ a Gaussian random
variable independent from all the other random variables. Then
\begin{align*}
\mathbb{C}\mathrm{ov}\left[\left(R_{t+\Delta t}-R_{t}\mid R_{t}\right),\left(W_{t+\Delta t}^{r}-W_{t}^{r}\mid R_{t}\right)\right] & =\mathbb{C}\mathrm{ov}\left[R_{t+\Delta t}-R_{t}\mid R_{t},\rho\left(W_{t+\Delta t}^{r}-W_{t}^{r}\right)+\sqrt{1-\rho^{2}}G\mid R_{t}\right]\\
 & =\mathbb{C}\mathrm{ov}\left[R_{t+\Delta t}-R_{t}\mid R_{t},\rho\left(W_{t+\Delta t}^{r}-W_{t}^{r}\right)\mid R_{t}\right]\\
 & =\frac{\rho}{\kappa}\left(1-e^{-\kappa\Delta t}\right).
\end{align*}
\begin{align*}
\mathbb{C}\mathrm{ov}\left[\left(I_{t+\Delta t}-I_{t}\mid R_{t}\right),\left(W_{t+\Delta t}^{r}-W_{t}^{r}\mid R_{t}\right)\right] & =\mathbb{C}\mathrm{ov}\left[I_{t+\Delta t}-I_{t}\mid R_{t},\rho\left(W_{t+\Delta t}^{r}-W_{t}^{r}\right)+\sqrt{1-\rho^{2}}G\mid R_{t}\right]\\
 & =\frac{\rho}{\kappa^{2}}\left(\kappa\Delta t+e^{-\kappa\Delta t}-1\right).
\end{align*}

\section{\label{sec:Appendix_B}Appendix: proof of Proposition \ref{prop:4}}

According to Proposition \ref{prop:3}, we can write
\[
\left(\begin{array}{c}
R_{t+\Delta t}-R_{t}\\
W_{t+\Delta t}^{S}-W_{t}^{S}
\end{array}\right)\mid R_{t}\sim\mathcal{N}\left(\left(\begin{array}{c}
\mu_{R}\\
\mu_{W}
\end{array}\right),\left(\begin{array}{cc}
\sigma_{R}^{2} & \sigma_{R,W}\\
\sigma_{R,W} & \sigma_{W}^{2}
\end{array}\right)\right),
\]
with
\[
\mu_{R}=R_{t}\left(e^{-\kappa\Delta t}-1\right),\ \mu_{W}=0,\ \sigma_{R}^{2}=\frac{1}{2\kappa}\left(1-e^{-2\kappa\Delta t}\right),\ \sigma_{W}^{2}=\Delta t,\ \text{and}\ \sigma_{R,W}=\frac{\rho}{\kappa}\left(1-e^{-\kappa\Delta t}\right).
\]
Now, let us define the linear coefficient of correlation between $R_{t+\Delta t}-R_{t}\mid R_{t}$
and $W_{t+\Delta t}^{S}-W_{t}^{S}\mid R_{t}$ as
\[
\rho_{W,R}=\frac{\sigma_{R,W}}{\sqrt{\sigma_{R}^{2}\sigma_{W}^{2}}}.
\]
A standard result about conditional Gaussian distributions (see e.g.
Theorem 3.5 in DasGupta \cite{dasgupta2011probability}) allows us
to compute the distribution of the Gaussian increment $W_{t+\Delta t}^{S}-W_{t}^{S}$
given the values $R_{t}$ and $R_{t+\Delta t}$ as follows: 
\[
\left(W_{t+\Delta t}^{S}-W_{t}^{S}\mid R_{t+\Delta t},R_{t}\right)\sim\mathcal{N}\left(\mu_{W\mid R},\sigma_{W\mid R}^{2}\right),
\]
with 
\begin{align*}
\mu_{W\mid R} & =\mu_{W}+\rho_{W,R}\sqrt{\frac{\sigma_{W}^{2}}{\sigma_{R}^{2}}}\left(R_{t+\Delta t}-\mu_{R}\right)\\
 & =\mu_{W}+\frac{\sigma_{R,W}}{\sigma_{R}^{2}}\left(R_{t+\Delta t}-\mu_{R}\right)\\
 & =0+\frac{\rho\frac{1-e^{-\kappa\Delta t}}{\kappa}}{\frac{1}{2\kappa}\left(1-e^{-2\kappa\Delta t}\right)}\left(R_{t+\Delta t}-R_{t}e^{-\kappa\Delta t}\right)\\
 & =2\rho\frac{1-e^{-\kappa\Delta t}}{1-e^{-2\kappa\Delta t}}\left(R_{t+\Delta t}-R_{t}e^{-\kappa\Delta t}\right),
\end{align*}
and 
\begin{align*}
\sigma_{W\mid R}^{2} & =\sigma_{W}^{2}\left(1-\rho_{W,R}^{2}\right)\\
 & =\sigma_{W}^{2}-\frac{\left(\sigma_{R,W}\right)^{2}}{\sigma_{R}^{2}}\\
 & =\Delta t-\frac{\left(\rho\frac{1-e^{-\kappa\Delta t}}{\kappa}\right)^{2}}{\frac{1}{2\kappa}\left(1-e^{-2\kappa\Delta t}\right)}\\
 & =\Delta t-2\frac{\rho^{2}}{\kappa}\frac{1-e^{-\kappa\Delta t}}{1+e^{-\kappa\Delta t}}.
\end{align*}

\subsubsection*{Acknowledgement}

This work was supported by the Departmental Strategic Plan (PSD) of
the University of Udine, \emph{Interdepartmental Project on Artificial
Intelligence (2020-2025).} 

\subsubsection*{Disclosure statement}

No potential conflict of interest was reported by the authors.

\bibliographystyle{plain}
\bibliography{bibliography}

\end{document}